\documentclass[a4paper,10pt,twoside]{cpc-hepnp}

\usepackage{multicol}
\usepackage{graphicx}
\usepackage{booktabs}
\usepackage{amssymb,bm,mathrsfs,bbm,amscd}
\usepackage[tbtags]{amsmath}
\usepackage{lastpage}
\usepackage{CJK}
\usepackage{color}
\usepackage{epstopdf}
\usepackage{epsfig}
\usepackage{ulem}
\usepackage{pdfsync}

\begin{document}
\begin{CJK*}{GB}{gbsn}

\title{Studying the potential of QQq at finite temperature in a holographic model}

\author{Xun Chen(³ÂÑ«)$^{1;1)}$\email{chenxunhep@qq.com}
\quad Bo Yu(Ó÷²©)$^{1}$
\quad Peng-Cheng Chu(³õÅô³Ì)$^{2,3;2)}$\email{kyois@126.com}
\quad Xiao-hua Li(ÀîС»ª)$^{1;3)}$\email{lixiaohuaphysics@126.com}
}
\maketitle

\address{%
$^1$ School of Nuclear Science and Technology, University of South China, Hengyang 421001, China\\
$^2$ The Research Center for Theoretical Physics, Science School, Qingdao University of Technology, Qingdao 266033, China\\
$^3$ The Research Center of Theoretical Physics, Qingdao University of Technology, Qingdao 266033, China \\
}

\begin{abstract}
Using gauge/gravity duality, we investigate the string breaking and dissolution of two heavy quarks coupled to a light quark at finite temperature. It is found that three configurations of QQq exist with the increase in separation distance for heavy quarks in the confined phase. Furthermore, string breaking occurs at the distance $L_{\rm{QQq}} = 1.27 \rm{fm}$($T = 0.1 \rm{GeV}$) for the decay mode $\rm{Q Q q \rightarrow Q q q+Q \bar{q}}$. In the deconfined phase, QQq melts at a certain distance then becomes free quarks. Finally, we compare the potential of QQq with that of $\rm{Q\bar{Q}}$ and it is found that  $\rm{Q\bar{Q}}$ is more stable than QQq at high temperature.
\end{abstract}


\begin{multicols}{2}
\section{Introduction}\label{sec-int}
With over 20 years of deveelopment, gauge/gravity duality has become a useful tool to deal with the QCD problem through gravitational theory.Quark-antiquark potential is  one of the hottest topics in holographic QCD , because heavy quarkonia are among the most sensitive probes used in the experimental study of quark-gluon plasma(QGP) and its properties. The holographic potential of the quark-antiquark pair was first recorded in Ref.\cite{Maldacena:1998im}. It was  found that the quark-antiquark potential exhibits a purely Coulombian\ (non-confining) behavior and agrees with a conformal gauge theory. Soon after, the potential at finite temperature was been discussed in \cite{Rey:1998bq,Brandhuber:1998bs}. The deformed $AdS_5$ model and Einstein-Maxwell-Dilation models were used to calculate the quark-antiquark potential in these studies\cite{Andreev:2006ct,Andreev:2006eh,Andreev:2006nw,He:2010bx,Colangelo:2010pe,DeWolfe:2010he,Li:2011hp,Fadafan:2011gm,Fadafan:2012qy,Cai:2012xh,Li:2012ay,Yang:2015aia,Zhang:2015faa,Fang:2015ytf,Ewerz:2016zsx,Chen:2017lsf,Arefeva:2018hyo,
Chen:2018vty,Bohra:2019ebj,Chen:2019rez,Zhou:2020ssi,Zhou:2021sdy,Chen:2020ath,Chen:2021gop}.

Moreover, the recent discovery of a doubly charmed baryon\ (DHB) $\Xi_{c c}+$ through LHCb experiments at CERN\cite{LHCb:2017iph,LHCb:2018pcs} has reinforced interest in the search for a theoretical description of doubly heavy baryons. Inside a DHB, there is a heavy diquark and light quark. Because the heavy quark in a DHB is almost near its mass shell, it is reasonable to expect the heavy quark limit to be applicable in this system\cite{Ma:2017nik}. Although lattice gauge theory remains a basic tool for studying nonperturbative phenomena in QCD, it has achieved limited results on QQq potentials to date \cite{Yamamoto:2008jz,Najjar:2009da}.

In recent years, the multi-quark potential from the holographic model has been discussed by Oleg Andreev in Ref.\cite{Andreev:2020xor,Andreev:2015iaa,Andreev:2015riv,Andreev:2019cbc,Andreev:2021bfg}. In this effective string model, heavy quarks are connected by string to a baryon vertex whose action is given by a five brane wrapped around an internal space, and the light quark is a tachyon field coupled to the worldsheet boundary. The main reason for pursuing this model is that its results on the quark-antiquark and three-quark potentials are consistent with lattice calculations and QCD phenomenology\cite{Andreev:2020xor,Andreev:2006ct,Andreev:2015iaa,Andreev:2015riv,Andreev:2019cbc,Andreev:2021bfg}. The technology we use to extract the QQq potential is same as at in lattice QCD\cite{Yamamoto:2008fm}. The QQq potential is extracted from the expectation value of the QQq Wilson loop ${\rm W_{QQq}}(R, T)$. The QQq Wilson loop is constructed from the heavy-quark trajectories and light-quark propagator. Hence, we investigate the QQq potential at finite temperature in this paper using gauge/gravity duality.

The remainder of this paper is organized as follows: We  establish the different configurations of string at finite temperature in Sec.2. Then, we numerically solve these configurations at different temperatures and provind a discussion in Sec.3. In Sec.4, we discuss string breaking for the confined phase. Finally, the summary and conclusion are given in Sec.5.

\section{Connected string configuration}\label{Connected}
First, we briefly review the holographic model used in the paper. Following Ref.~\cite{Andreev:2015riv}, this background metric at finite temperature is
\begin{equation}
d \mathbf{s}^{2}=\mathrm{e}^{s r^{2}} \frac{R^{2}}{r^{2}}\left(f(r)d t^{2}+d \vec{x}^{2}+ f^{-1}(r) d r^{2}\right)+\mathrm{e}^{-s r^{2}} g_{a b}^{(5)} d \omega^{a} d \omega^{b}.
\end{equation}
This model parameterized by $s$ is a one-parameter deformation of Euclidean $AdS_5$ space with a constant radius $R$, and a  five-dimensional compact space (sphere) $\textbf{X}$, whose coordinates are $\omega^a$ and $f(r)$ ,is a blackening factor. The Nambu-Goto action of a string is expressed as
\begin{equation}
S=\frac{1}{2 \pi \alpha^{\prime}} \int_{0}^{1} d \sigma \int_{0}^{T} d \tau \sqrt{\gamma},
\end{equation}
where $\gamma$ is an induced metric on the string world-sheet (with a Euclidean signature). For $AdS_5$ space, $f(r) = 1- \frac{r^4}{r_h^4}$ with the boundary conditions $f(0) = 1$ and $f(r_h) = 0$. $r_h$ is the position of the black hole(brane). The Hawking temperature identified with the temperature of a dual gauge theory can be defined as $T=\frac{1}{4 \pi}\left|\partial_{r} f\right|_{r=r_{h}}$. The motivations for this metric have been clarified in Ref.~\cite{Andreev:2015riv}. However, such a deformed $AdS_5$ metric leads to linear Regge-like spectra for mesons\cite{Karch:2006pv,Andreev:2006vy} and the Cornell potential of a quark-antiquark \cite{Andreev:2006ct}. The deformed metric
satisfies the thermodynamics of lattice\cite{Andreev:2007zv}.

Subsequently, we introduce the baryon vertex. According to the AdS/CFT correspondence, this is a five brane\cite{Witten:1998xy}. At leading order in $\alpha'$, the brane action is $S_{vert} =\mathcal{T}_5 \int d^6 \xi \sqrt{\gamma^{(6)}}$, where $\mathcal{T}_5$ is the brane tension, and $\xi^{i}$ are the world-volume coordinates. Because the brane is wrapped around the internal space, it appears point-like in $AdS_5$. We choose a static gauge $\xi^0 = t$ and $\xi^a = \theta^a$ with $\theta^a$ coordinates on $\textbf{X}$. Thus, the action is

\begin{equation}
S_{\mathrm{vert}}=\tau_{v} \int d t \frac{\mathrm{e}^{-2 s r^{2}}}{r} \sqrt{f(r)},
\end{equation}
where $\tau_{v}$ is a dimensionless parameter defined by $\tau_{v}=\mathcal{T}_{5} R \operatorname{vol}(\mathbf{X})$ and $\operatorname{vol}(\mathbf{X})$ is a volume of $\textbf{X}$.

Finally, we consider that the light quark at string endpoints is a tachyon field, which couples to the world-sheet boundary via  $S_{\mathrm{q}}=\int d \tau e \mathrm{~T}$; this is the usual sigma-model action  for a string propagating in the tachyon background\cite{Andreev:2020pqy}. The integral is over a world-sheet boundary parameterized by $\tau$ ,and $e$ is a boundary metric. We consider a constant background $\rm{T}(x,r) = \rm{T_0}$ and worldsheets whose boundaries are lines in the $t$ direction. Thus, the action can be written as
\begin{equation}
S_{\mathrm{q}}=\mathrm{m} \int d t \frac{\mathrm{e}^{\frac{s}{2} r^{2}}}{r} \sqrt{f(r)},
\end{equation}
where $\mathrm{m}=R \mathrm{~T}_{0}$. This is the action of a point particle of mass $\mathrm{T_0}$ at rest. We choose the model parameters as follows: $\mathbf{g}=\frac{R^{2}}{2 \pi \alpha^{\prime}}, \mathrm{k}=\frac{\tau_{v}}{3 \mathbf{g}}$, and $\mathrm{n}=\frac{\mathrm{m}}{\mathbf{g}}$.

\subsection{Small $L$}

\begin{center}
	\includegraphics[width=8cm]{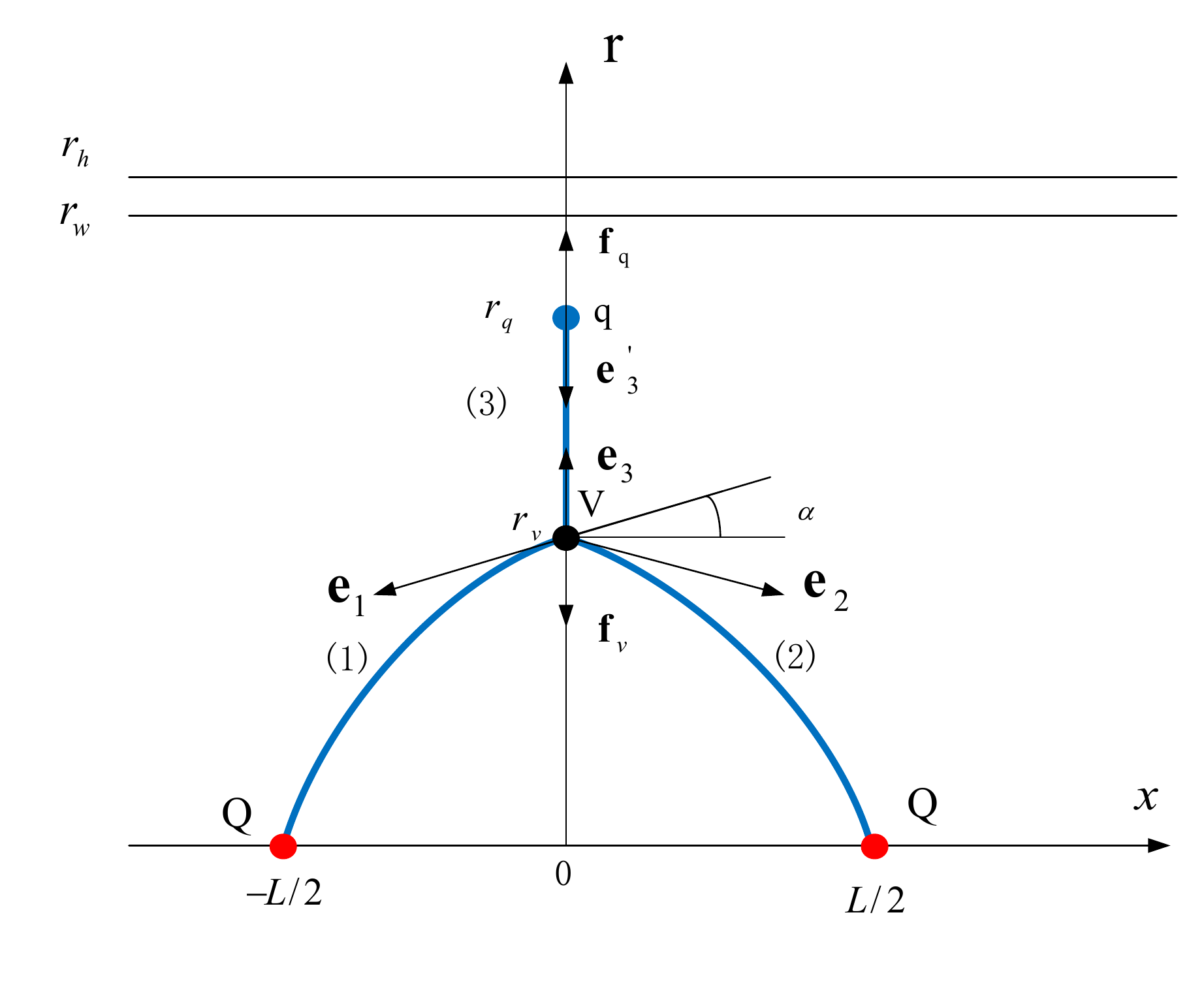}
	\figcaption{\label{SL} Static string configuration at  a small separation distance of a heavy-quark pair. The heavy quarks Q are placed on the x-axis, while the light quark q and baryon vertex V are on the r-axis, at $r = r_q$ and $r = r_v$ respectively. The quarks and baryon vertex are connected by the blue string. The force exerted on the vertex and light quark are depicted by the black arrows. $r_h$ is the position of the black-hole horizon. $r_w$ is the position of a soft wall in the confined phase.}
\end{center}

The configuration of QQq is plotted in Fig.\ref{SL}. The total action is the sum of the Nambu-Goto actions plus the actions for the vertex and background scalar.
\begin{equation}
S=\sum_{i=1}^{3} S_{\mathrm{NG}}^{(i)}+S_{\mathrm{vert}}+S_{\mathrm{q}}.
\end{equation}
If we choose the static gauge $\xi^{1}=t$ and $\xi^{2}=r$,  the boundary conditions for $x(r)$ become
\begin{equation}
x^{(1)}(0)=-\frac{L}{2} , \quad x^{(2)}(0)=\frac{L}{2}, \quad x^{(i)}\left(r_{v}\right)=x^{(3)}\left(r_{q}\right)=0.
\end{equation}
The action can now be written as
\begin{equation}\label{totalaction1}
\begin{aligned}
S&=\mathbf{g} T \Big(2 \int_{0}^{r_{v}} \frac{d r}{r^{2}}  \mathrm{e}^{s r^{2}} \sqrt{1+f(r)\left(\partial_{r} x\right)^{2}}\\&+\int_{r_{v}}^{r_{q}} \frac{d r}{r^{2}} \mathrm{e}^{\mathrm{s} r^{2}} \sqrt{1+f(r)\left(\partial_{r} x\right)^{2}}+3 \mathrm{k} \frac{\mathrm{e}^{-2 \mathrm{~s} r_{v}^{2}}}{r_{v}}\sqrt{f(r_v)}\\&+\mathrm{n} \frac{\mathrm{e}^{\frac{1}{2} \mathrm{~s} r_{q}^{2}}}{r_{q}}\sqrt{f(r_q)}\Big),
\end{aligned}
\end{equation}
where $\partial_{r} x=\frac{\partial x}{\partial r}$ and $T=\int_{0}^{T} d t$. We consider the first term in~(\ref{totalaction1}), which corresponds to string (1) and (2) in Fig.~\ref{SL}. The equation of motion(EoM) for $x(r)$ can be obtained from the  Euler-Lagrange equation. Thus, we have

\begin{equation}
\mathcal{I}=\frac{w(r) f(r) \partial_{r} x}{\sqrt{1+f(r)\left(\partial_{r} x\right)^{2}}}, \quad w(r)=\frac{\mathrm{e}^{\mathrm{s} r^{2}}}{r^{2}}.
\end{equation}
$\mathcal{I}$ is a constant. At $r_v$, we have $\left.\partial_{r} x\right|_{r=r_{v}}=\cot \alpha$ with $\alpha>0$ and
\begin{equation}
\mathcal{I}=\frac{w(r_v) f(r_v) cot \alpha}{\sqrt{1+f(r)\left(cot \alpha\right)^{2}}}.
\end{equation}
$\partial_{r} x$ can be solved as

\begin{equation}\label{xp1}
\partial_{r} x = \sqrt{\frac{\omega\left(r_{v}\right)^{2} f\left(r_{v}\right)^{2}}{\left(f\left(r_{\nu}\right)+\tan ^{2} \alpha\right) \omega(r)^{2} f(r)^{2}-f(r) w\left(r_{v}\right)^{2} f(r_v)^{2}}}.
\end{equation}
Using ~(\ref{xp1}), the integral over [$0,r_v$] of $dr$ is

\begin{equation}\label{distance1}
L=2 \int_{0}^{r_{v}} \frac{dx}{dr} dr.
\end{equation}
$L$ is a function of $r_v$, $\alpha$ ,and $r_h$(or equally, temperature). The energy of string (1) can be found from the first term of ~(\ref{totalaction1}):

\begin{equation}
E_R=\frac{S}{T}=\mathbf{g} \int_{0}^{r_{v}}\frac{d r}{r^{2}}  \mathrm{e}^{\mathbf{s} r^{2}} \sqrt{1+f(r)\left(\partial_{r} x\right)^{2}}.
\end{equation}
Subtracting the divergent term $\mathbf{g} \int_{0}^{\infty}dr\frac{1}{r^2}$, we have the regularized energy:

\begin{equation}\label{energy1}
E_{1}=\frac{S}{T}=\mathbf{g} \int_{0}^{r_{v}}(\frac{1}{r^{2}}  \mathrm{e}^{\mathbf{s} r^{2}} \sqrt{1+f(r)\left(\partial_{r} x\right)^{2}}-\frac{1}{r^2})d r - \frac{\mathbf{g}}{r_v} + c.
\end{equation}
Here $c$ is a normalization constant. because string (2) produces the same results, we move to string (3) whose action is given by the second term in ~(\ref{totalaction1}). This string is a straight string stretched between the vertex and light. The energy can be calculated as

\begin{equation}
E_{2}=\mathbf{g} \int_{r_{v}}^{r_{q}} \frac{d r}{r^{2}} \mathrm{e}^{\mathrm{s} r^{2}}.
\end{equation}

Now, we can present the energy of the configuration. From ~(\ref{totalaction1}), it follows that
\begin{equation}\label{freeenergy1}
\begin{aligned}
E_{Q Q_{q}} &= \mathbf{g} \Big(2\int_{0}^{r_{v}}(\frac{1}{r^{2}}  \mathrm{e}^{\mathbf{s} r^{2}} \sqrt{1+f(r)\left(\partial_{r} x\right)^{2}}-\frac{1}{r^2})d r - \frac{2}{r_v}\\
&+ \int_{r_{v}}^{r_{q}} \frac{d r}{r^{2}} \mathrm{e}^{\mathrm{s} r^{2}}+  \mathrm{n} \frac{\mathrm{e}^{\frac{1}{2} s r_q^2}}{r_q}\sqrt{f(r_q)} +3 \mathrm{k} \frac{\mathrm{e}^{-2 s r_v^2}}{r_v}\sqrt{f(r_q)}\Big)\\
&+2 c.
\end{aligned}
\end{equation}

The energy is also a function of $r_v$, $\alpha$, and $r_h$. There are two steps to be compeleted: The first is to determine the position of the light quark, which is a function of temperature,and the second is to determine $\alpha$. To achieve this goal, the net forces exerted on the light quark and vertex vanish .
We first write the force balance equation of  the  light quark as
\begin{equation}
\mathbf{f}_{q}+\mathbf{e}_{3}^{\prime}=0.
\end{equation}
By varying the action with respect to $r_q$, it is found that $\mathbf{f}_{q}=\left(0,-\mathbf{g} n \partial_{r_{q}} (\frac{\mathrm{e}^{\frac{1}{2} \mathrm{sr}_{q}^{2}}}{r_{q}}\sqrt{f(r_q)})\right)$ and $\mathbf{e}_{3}^{\prime}=\mathbf{g} w\left(r_{q}\right)(0,-1)$. Hence, the force balance equation becomes
\begin{equation}\label{lightquarkforce}
2e^{\frac{r_q^{2} s}{2}}\sqrt{f(r_q)}+2n\left(-1+r_q^{2} s\right) f(r_q)+n r_q f^{\prime}(r_q)= 0.
\end{equation}
Note that $r_q$
(the position of the light quark) is only a function of $r_h$. At fixed temperature, $r_q$ can be fixed via the  above equation. The force balance equation on the vertex is
\begin{equation}\label{force balance}
\mathbf{e}_{1}+\mathbf{e}_{2}+\mathbf{e}_{3}+\mathbf{f}_{v}=0.
\end{equation}
Here $\mathbf{e}_{i}$ is the string tension which can be calculated in the same manner as in Ref.\cite{Andreev:2021bfg}. As a result, the force on the vertex is $\mathbf{f}_{v}=\left(0,-3 \mathbf{g} k \partial_{r_{v}} (\frac{\mathrm{e}^{-2 s r_{v}^{2}}}{r_{v}}\sqrt{f(r_v)})\right)$, and the string tensions are $\mathbf{e}_{1}=\mathbf{g} w\left(r_{v}\right)(-\frac{f(r_{v})}{\sqrt{tan^2 \alpha + f(r_{v})}},-\frac{1}{\sqrt{f(r_{v}) cot^2 \alpha + 1}})$, $\mathbf{e}_{2}=\mathbf{g} w\left(r_{v}\right)(\frac{f(r_{v})}{\sqrt{tan^2 \alpha + f(r_{v})}},-\frac{1}{\sqrt{f(r_{v}) cot^2 \alpha + 1}})$, $\mathbf{e}_{3}=\mathbf{g} w\left(r_{v}\right)(0,1)$. The non-trivial component of the force balance equation is

\begin{equation}\label{force balance equation}
2 e^{3 s r_v^{2}}(1-\frac{2}{1+cot\alpha^2 f(r_v)})+6\left(k+4 k s r_v^{2} \right)\sqrt{f(r_{v})}-\frac{3 k r_v f^{\prime}(r_v)}{\sqrt{f(r_v)}}=0.
\end{equation}
This equation provides the relationship between  $r_v$ and $\alpha$ when the temperature is fixed.  With ~(\ref{freeenergy1}) and ~(\ref{lightquarkforce}), we can numerically solve the energy at small $L$.

\subsection{Intermediate $L$}
\begin{center}
	\includegraphics[width=8cm]{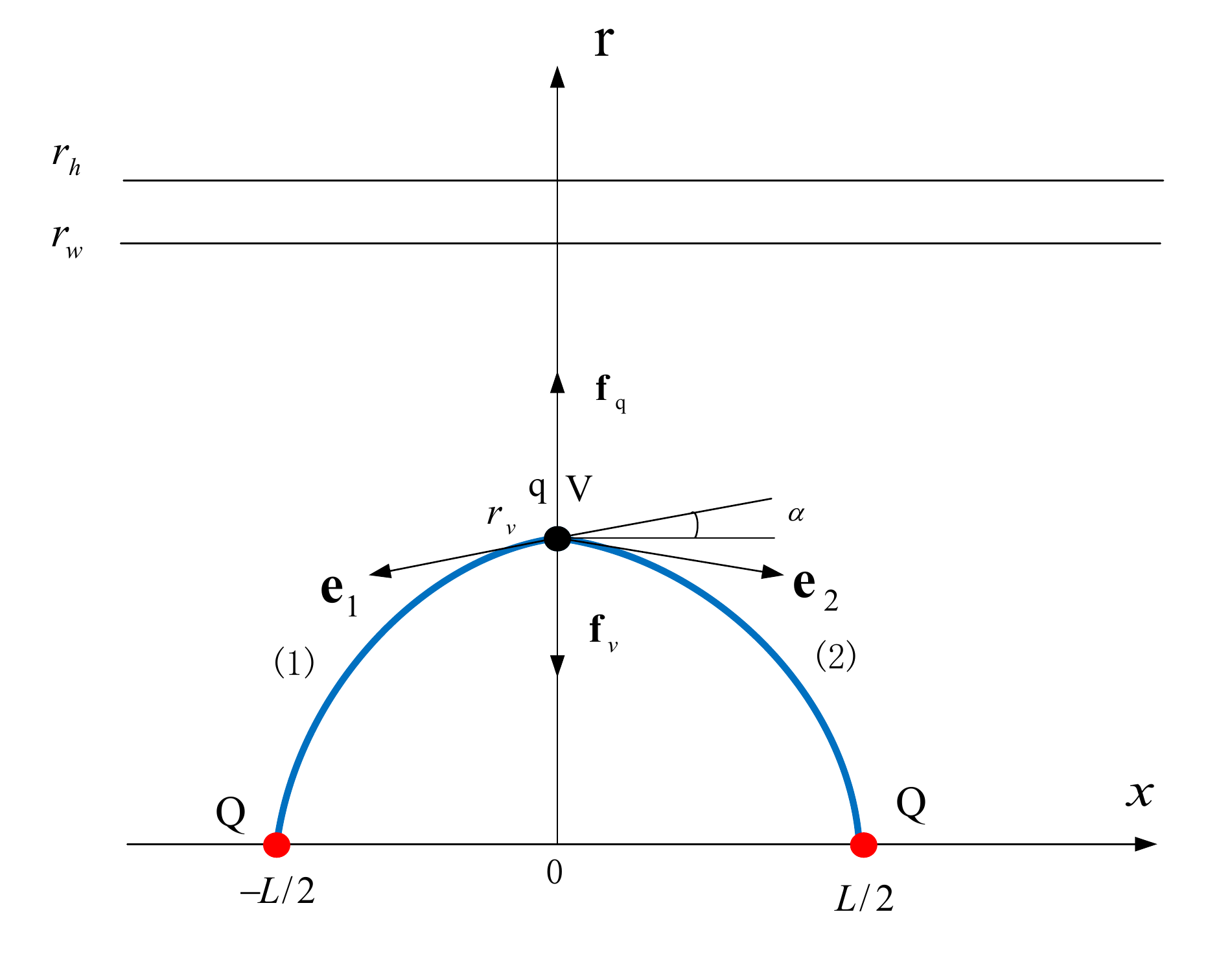}
	\figcaption{\label{IL} Static string configuration at an intermediate separation distance of a heavy-quark pair. The heavy quarks Q are placed on the x-axis, while the light quark q and baryon vertex V are at the same point $r_v$ on the r-axis. The forces exerted on the point are depicted by the black arrows. $r_h$ is the position of the black-hole horizon. $r_w$ is the position of a soft wall in the confined phase.}
\end{center}
The total action of the configuration plotted in Fig.\ref{IL}is expressed as
\begin{equation}\label{totalaction2}
S=\sum_{i=1}^{2} S_{\mathrm{NG}}^{(i)}+S_{\mathrm{vert}}+S_{\mathrm{q}}.
\end{equation}
We still choose the same static gauge as before. The boundary conditions then take the form
\begin{equation}
x^{(1)}(0)=-\frac{L}{2}, \quad x^{(2)}(0)=\frac{L}{2} , \quad x^{(i)}\left(r_{v}\right)=0.
\end{equation}
In this configuration, the expressions~(\ref{distance1}) and ~(\ref{energy1}) still hold. Naturally, we can express the total energy of the string without the contribution from string (3) as follows:
\begin{equation}\label{freeenergy2}
\begin{aligned}
E_{Q Q_{q}}=\mathbf{g} (2\int_{0}^{r_{v}}(\frac{1}{r^{2}}  \mathrm{e}^{\mathbf{s} r^{2}} \sqrt{1+f(r)\left(\partial_{r} x\right)^{2}}-\frac{1}{r^2})d r - \frac{2}{r_v}\\ +  \mathrm{n} \frac{\mathrm{e}^{\frac{1}{2} s r_v^2}}{r_v} \sqrt{f(r_v)} +3 \mathrm{k} \frac{\mathrm{e}^{-2 s r_v^2}}{r_v}\sqrt{f(r_v)})+2 c.
\end{aligned}
\end{equation}
The force balance equation at the point $r = r_v$ is
\begin{equation}\label{forcebalance2}
\mathbf{e}_{1}+\mathbf{e}_{2}+\mathbf{f}_{v}+\mathbf{f}_{q}=0.
\end{equation}
Each force is given by
$$\mathbf{f}_{q}=\left(0,-n \mathbf{g}  \partial_{r_{q}}\left(\frac{\mathrm{e}^{\frac{1}{2} s r_{q}^{2}}}{r_{q}} \sqrt{f\left(r_{q}\right)}\right)\right),$$

$$\mathbf{f}_{v}=\left(0,-3 \mathbf{g}  k \partial_{r_{v}}\left(\frac{\mathrm{e}^{-2 s r_{v}^{2}}}{r_{v}} \sqrt{f\left(r_{v}\right)}\right)\right),$$

$$\mathbf{e}_{1}=\mathbf{g} w\left(r_{v}\right)(-\frac{f(r_{v})}{\sqrt{tan^2 \alpha + f(r_{v})}},-\frac{1}{\sqrt{f(r_{v}) cot^2 \alpha + 1}}),$$

$$\mathbf{e}_{2}=\mathbf{g} w\left(r_{v}\right)(\frac{f(r_{v})}{\sqrt{tan^2 \alpha + f(r_{v})}},-\frac{1}{\sqrt{f(r_{v}) cot^2 \alpha + 1}}),$$
where $r_{q} = r_{v}$. The force balance equation leads to
\begin{equation}\label{force balance equation2}
\begin{aligned}
2(-e^{\frac{5 r_v^2 s}{2}} n (-1 + r_v^2 s)+3 k (1+4r_v^2 s))\sqrt{f(r_v)}\\-\frac{4e^{3r_v^2 s}}{\sqrt{1+cot\alpha^2 f(r_v)}}-\frac{(3k+e^{\frac{5r_v^2 s}{2}})r_v f'(r_v)}{\sqrt{f(r_v)}}=0.
\end{aligned}
\end{equation}

\subsection{Large $L$}
\begin{center}
	\includegraphics[width=8cm]{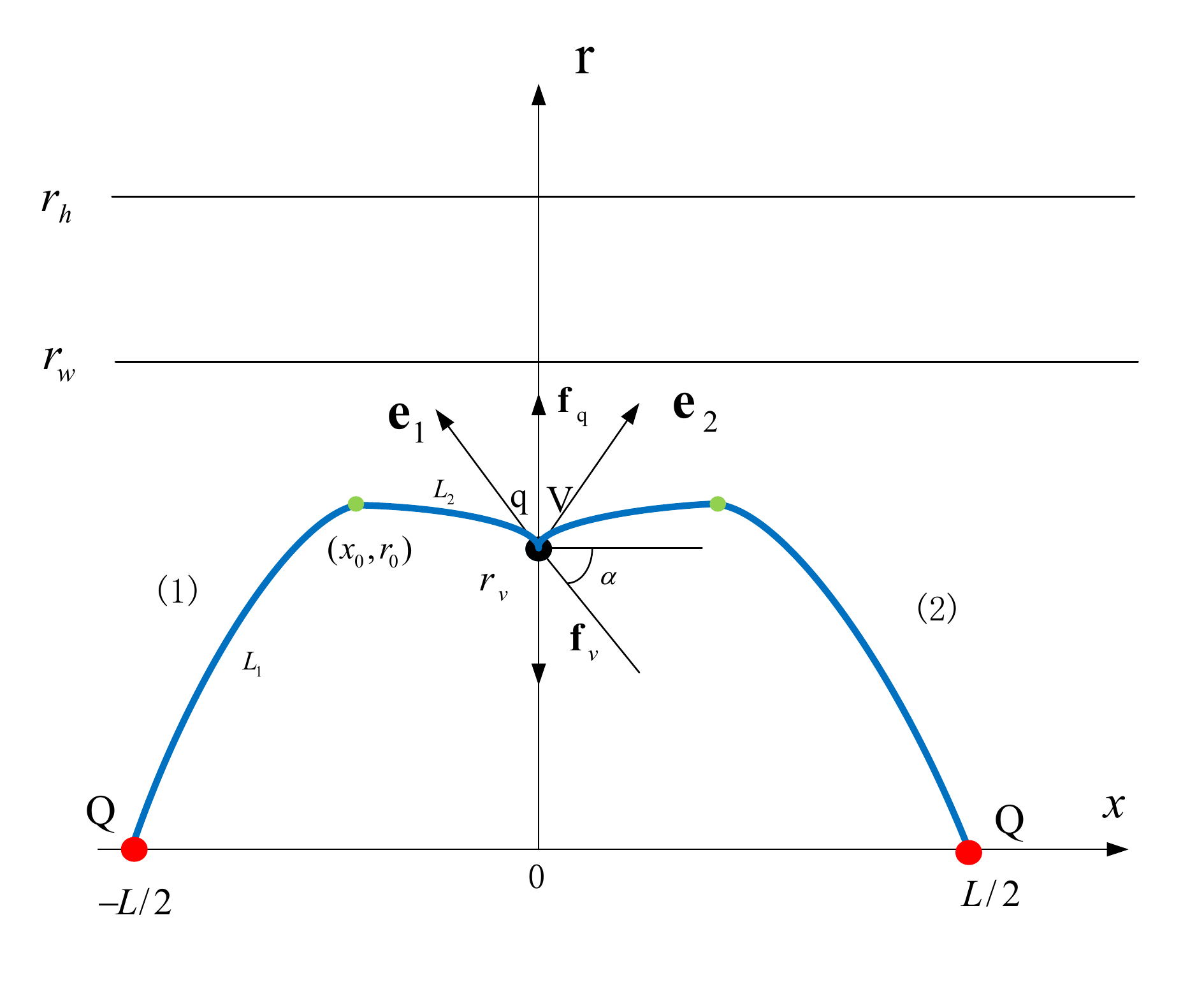}
	\figcaption{\label{LL} Static string configuration at A large separation distance of a heavy-quark pair. The heavy quarks Q are placed on the x-axis, while the light quark q and baryon vertex V are at the same point $r_v$ on the r-axis. The forces  exerted on the point are depicted by the black arrows. There is a turning point located at $(x_0, r_0)$ for string (1). $r_w$ is the position of a soft wall in the confined phase. $r_h$ is the position of the black-hole horizon.}
\end{center}
The total action of the configuration plotted in Fig.\ref{LL}is the same as Eq.~(\ref{totalaction2}). However, it is  convenient to choose another static gauge $\xi^{1}=t$ and $\xi^{2}=x$ here. Then, the boundary conditions are
\begin{equation}
r^{(1)}(-L / 2)=r^{(2)}(L / 2)=0, \quad r^{(i)}(0)=r_{v} .
\end{equation}
The total action becomes
\begin{equation}\label{Stotal3}
\begin{aligned}
S&=\mathbf{g} T\Big(\int_{-L / 2}^{0} d x w(r) \sqrt{f(r)+\left(\partial_{x} r\right)^{2}}+\int_{0}^{L / 2} d x w(r) \sqrt{f(r)+\left(\partial_{x} r\right)^{2}}\\
&+3 \mathrm{k} \frac{\mathrm{e}^{-2 s r_{v}^{2}}}{r_{v}}\sqrt{f(r_v)}+\mathrm{n} \frac{\mathrm{e}^{\frac{1}{2} \mathrm{~s} r_{v}^{2}}}{r_{v}}\sqrt{f(r_v)}\Big).
\end{aligned}
\end{equation}
We consider string (1), whose action is given by the first term in~(\ref{Stotal3}). The first integral has the form 
\begin{equation}
\mathcal{I}=\frac{w(r)f(r)}{\sqrt{f(r)+\left(\partial_{x} r\right)^{2}}}.
\end{equation}
At points $r_0$ and $r_v$, we have
\begin{equation}\label{first1}
\frac{w(r)f(r)}{\sqrt{f(r)+\left(\partial_{x} r\right)^{2}}}=w(r_0)\sqrt{f(r_0)},
\end{equation}

\begin{equation}\label{first2}
\frac{w(r_v)f(r_v)}{\sqrt{f(r_v)+ tan\alpha^2}}=w(r_0)\sqrt{f(r_0)}.
\end{equation}
The relationship between $r_0$, $r_v$, and $\alpha$ can be obtained from Eq.~(\ref{first2}). Using Eq.~(\ref{first1}), the separation distance and energy can be subsequently obtained. As before, $\partial_x r$ can be solved as
\begin{equation}\label{first}
\partial_x r =\sqrt{\frac{w(r)^2f(r)^2f(r_0) - f(r)w(r_0)^2 f(r_0)^2}{w(r_0)^2 f(r_0)^2}}.
\end{equation}
The separation distance consists of two parts:
\begin{equation}\label{distance2}
\begin{aligned}
L &= 2(L_1 + L_2) = 2(\int_0^{r_0} \frac{1}{r^{\prime}} dr + \int_{r_v}^{r_0} \frac{1}{r^{\prime}} dr)\\
&=2(\int_0^{r_0} \sqrt{\frac{w(r_0)^2 f(r_0)^2}{w(r)^2f(r)^2f(r_0) - f(r)w(r_0)^2f(r_0)^2}} dr \\
&+ \int_{r_v}^{r_0} \sqrt{\frac{w(r_0)^2 f(r_0)^2}{w(r)^2f(r)^2f(r_0) - f(r)w(r_0)^2f(r_0)^2}} dr).
\end{aligned}
\end{equation}
The energy of string (1) can be obtained by summing two parts:
\begin{equation}
\begin{aligned}
E_R &= E_{R_1} + E_{R_2}\\
&=\mathbf{g} \int_0^{r_0} w(r)\sqrt{1 + f(r) x^{\prime 2}} dr + \mathbf{g} \int_{r_v}^{r_0} w(r)\sqrt{1 + f(r) x^{\prime 2}} dr \\
&=\mathbf{g} \int_0^{r_0} w(r) \sqrt{\frac{w(r)^2 f(r)^2 f(r_0)}{w(r)^2f(r)^2f(r_0) - f(r) w(r_0)^2 f(r_0)^2}} dr \\
&+ \mathbf{g} \int_{r_v}^{r_0} w(r) \sqrt{\frac{w(r)^2 f(r)^2 f(r_0)}{w(r)^2f(r)^2f(r_0) - f(r) w(r_0)^2 f(r_0)^2}} dr .
\end{aligned}
\end{equation}
After subtracting the divergent term, the renormalized energy of string (1) is
\begin{equation}
\begin{aligned}
E &=\mathbf{g} \int_{r_v}^{r_0}  (w(r) \sqrt{\frac{w(r)^2 f(r)^2 f(r_0)}{w(r)^2f(r)^2f(r_0) - f(r) w(r_0)^2 f(r_0)^2}}) dr \\&+\mathbf{g} \int_0^{r_0} ( w(r) \sqrt{\frac{w(r)^2 f(r)^2 f(r_0)}{w(r)^2f(r)^2f(r_0) - f(r) w(r_0)^2 f(r_0)^2}}\\
 &- \frac{1}{r^2}) dr - \frac{1}{r_0}+ 2c
\end{aligned}
\end{equation}
The calculation of string (2) has the same procedure as before and gives the same expressions for $L$ and $E$. Then, the total energy of the configuration can be written as
\begin{equation}
\begin{aligned}
E_{QQq} &= \Big{(}  2\int_{r_v}^{r_0} w(r) \sqrt{\frac{w(r)^2 f(r)^2 f(r_0)}{w(r)^2f(r)^2f(r_0) - f(r) w(r_0)^2 f(r_0)^2}}\\ dr+& 2\int_0^{r_0} (w(r) \sqrt{\frac{w(r)^2 f(r)^2 f(r_0)}{w(r)^2f(r)^2f(r_0) - f(r) w(r_0)^2 f(r_0)^2}} \\- \frac{1}{r^2}) dr
& - \frac{1}{r_0}+ 3 \mathrm{k} \frac{\mathrm{e}^{-2 \mathrm{~s} r_{v}^{2}}}{r_{v}}+\mathrm{n} \frac{\mathrm{e}^{\frac{1}{2} \mathbf{s} r_{v}^{2}}}{r_{v}}\Big)\mathbf{g} + 2c.
\end{aligned}
\end{equation}
The force balance equation is the same as Eq.~(\ref{forcebalance2}). Each force is similarly given in the previous section. With Eq.~(\ref{first2}) and (\ref{force balance equation2}), we can numerically solve  $L$ and $E_{QQq}$.

Except the symmetric case, the light quark can be in a position far from the r-axis. The non-symmetric configuration has been discussed in Ref.\cite{Andreev:2020xor}. which, as noted in Ref.\cite{Andreev:2020xor}, this string configuration is not energetically favorable. In this paper, we mainly focus on the symmetric configuration.

\section{Numerical results and discussion}\label{Numerical}

The system will change from the confined to deconfined phase with increasing temperature. The procedure for determining the melting temperature of QQq is similar to that of $\rm Q\bar{Q}$. We can judge the meting temperature from the behavior of the potential energy. In the confined phase, if we do not consider  string breaking, the potential will rise linearly forever. When changing from  a low to high temperature, the behavior of the potential will have an endpoint as shown in the Fig.\ref{0145EL}. Besides the potential, we can also judge the melting temperature from the behavior of  the separation distance. Fourther discussions can be found in \cite{Andreev:2006nw,Chen:2017lsf,Yang:2015aia,Colangelo:2010pe}.

In this section, we investigate the effect of different temperatures on the  QQq potential. The configurations of QQq change with temperature. With an increase in temperature,  QQq  melts. Because we want to approach the lattice at vanishing temperature, all the parameters are fixed as follows: $s = 0.42GeV^2$, $\mathbf{g} = 0.176$, $n = 3.057$, $k = -\frac{1}{4}e^{\frac{1}{4}}$, and $c = 0.623GeV$\cite{Andreev:2020xor}.

\subsection{$T=0.1GeV$}
First, we investigate the configurations of QQq at the low temperature $T=0.1GeV$. At this temperature, the system is in the confined phase with a soft wall below the black-hole horizon and the string and light quark can not exceed this  wall\cite{Andreev:2006nw,Colangelo:2010pe,Chen:2017lsf,Yang:2015aia}.

\subsubsection{small $L$}
\begin{figure*}
	\centering
	\includegraphics[width=15cm]{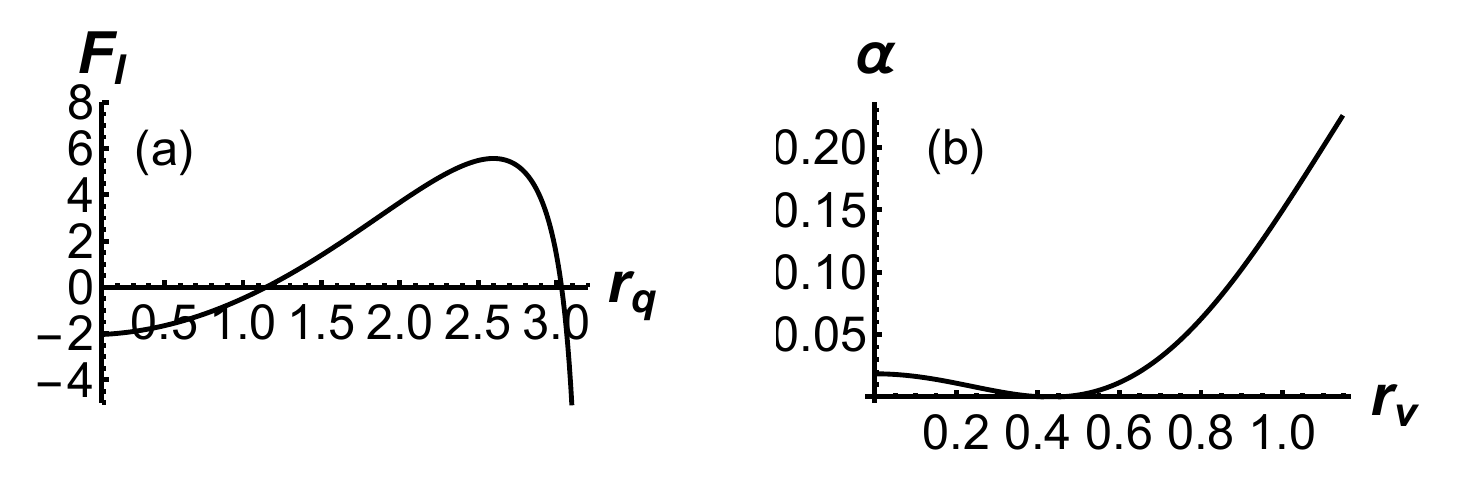}
	\caption{\label{01light} (a) Force balance equation of the light quark as a function of $r_q$. (b)$\alpha$ as a function of $r_v$. The unit for $r_q$ and $r_v$ is $\rm{GeV^{-1}}$.}
\end{figure*}

\begin{figure*}
	\centering
	\includegraphics[width=8cm]{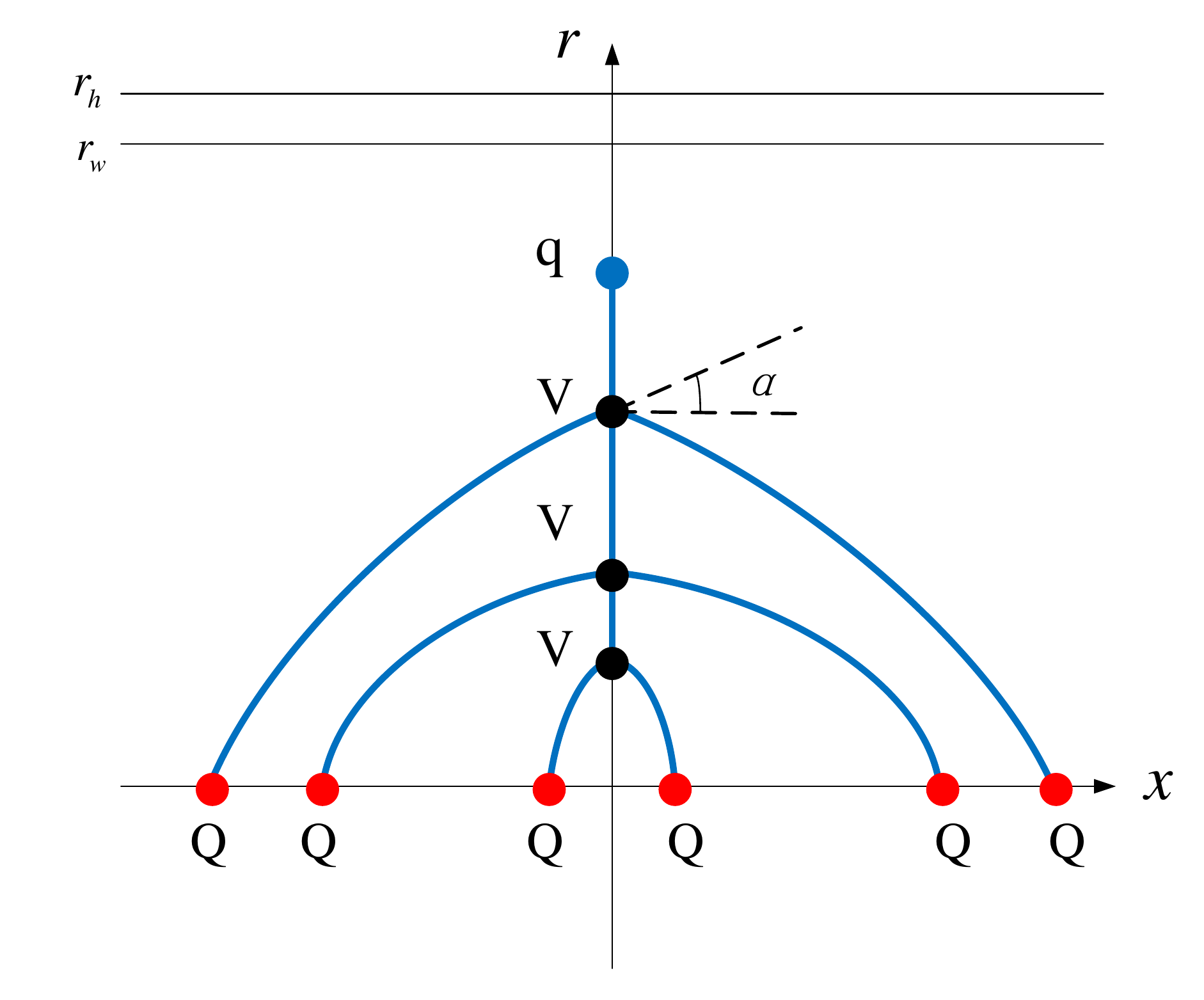}
	\caption{\label{01Sketch1} Schematic diagram of the string configuration with  increasing of $r_v$ for small $L$. $\alpha$ is always positive.}
\end{figure*}

From~(\ref{lightquarkforce}), we can determine the position of the light quark at fixed temperature. The force balance equation of the  light quark $F_l(r_q)$ as a function of $r_q$ is presented in Fig.~\ref{01light}(a).where two solutionsare observed. However, one solution beyond the soft wall is unphysical. At temperature $T=0.1GeV$, $r_q = 1.15 GeV^{-1}$ is a solution of the force balance equation. Then, $\alpha$ can be numerically solved from Eq.~(\ref{force balance equation}). The force balance equation for the vertex can  give the relationship between the angle $\alpha$ and $r_v$ as shown in Fig.~\ref{01light}(b). In the range  $0 \leq r_v \leq r_q$, we find that $\alpha$ is not a monotone function of $r_v$;it first decreases and then increases with an increase in $r_v$. A schematic diagram is shown in Fig.~\ref{01Sketch1}.

\begin{figure*}
	\centering
	\includegraphics[width=14cm]{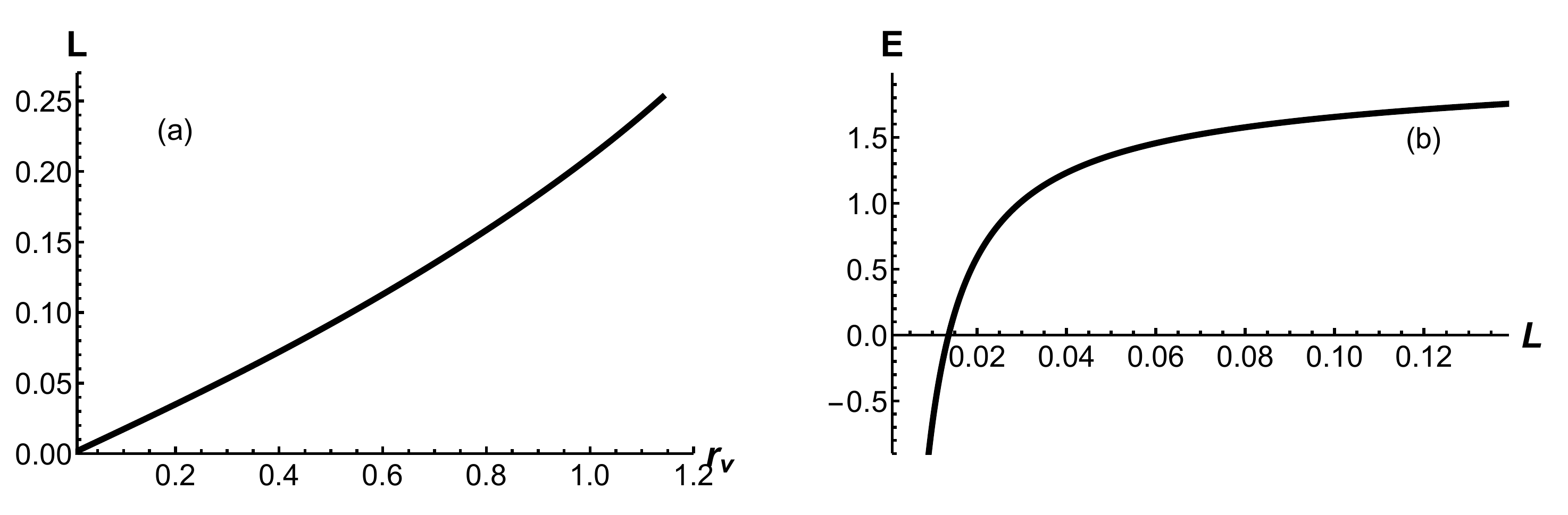}
	\caption{\label{01SLE} (a)Separation distance $L$ as a function of $r_v$. (b)Energy $E$ as a function of separation distance $L$. The unit of $E$ is $\rm{GeV}$, the unit of $L$ is $\rm{fm}$ and that of  $r_v$ is $\rm{GeV^{-1}}$.}
\end{figure*}

Furthermore, the separation distance of a heavy-quark pair which can be calculated using Eq.~(\ref{distance1}) is shown in Fig.~{\ref{01SLE}} (a). $L$ is a monotonously increasing function of $r_v$. At $r_v = r_q$, there is a cutoff. Beyond $r_q$, the configuration will turn into the second case. The corresponding energy can be calculated using Eq.~(\ref{freeenergy1}), which is shown in Fig.~\ref{01SLE} (b). It is found that the energy is a Coulomb potential at small separation distances and a linear potential at large distances.

\subsubsection{Intermediate $L$}
In the second configuration, $r_v$ ranges from $r_q$ to a certain position where $\alpha$ vanishes, as shown in Fig.~\ref{alpha1}. A schematic diagram for this case is shown in Fig.~\ref{01Sketch2}. At the beginning, $\alpha \approx 0.22$. When we increase $r_v$, $\alpha$ slowly tends to zero.

\begin{center}
	\includegraphics[width=8cm]{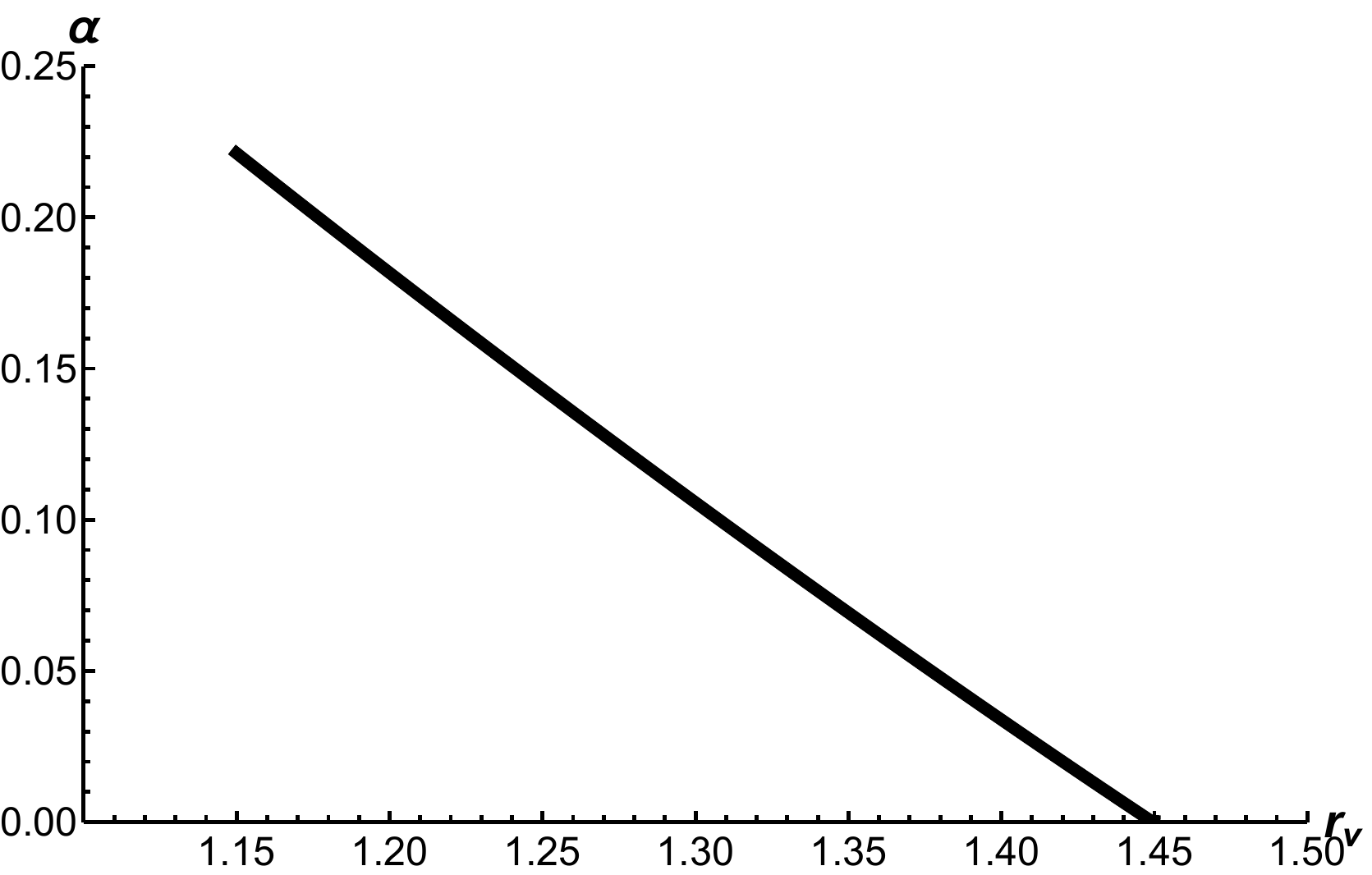}
	\figcaption{\label{alpha1}  $\alpha$ as a function of $r_v$. The unit of  $r_v$ is $\rm{GeV^{-1}}$.}
\end{center}

\begin{figure*}
	\centering
	\includegraphics[width=14cm]{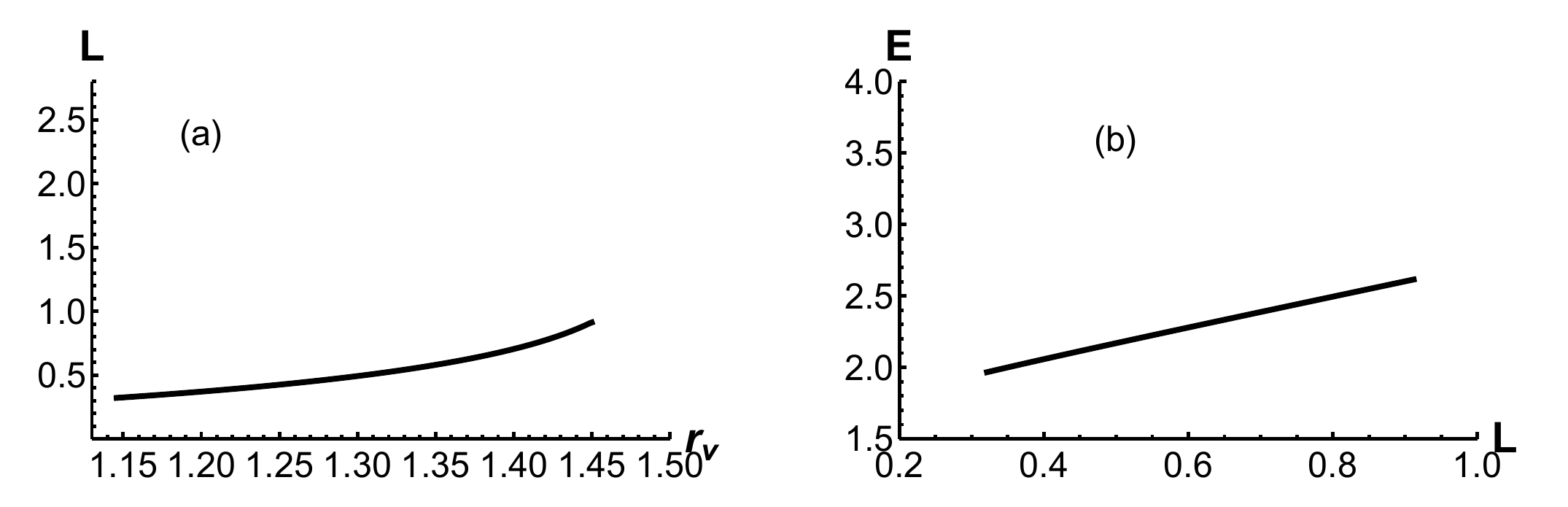}
	\caption{\label{01ILE} (a)Separation distance $L$ as a function of $r_v$. (b)The energy $E$ as a function of $L$. The unit of $E$ is $\rm{GeV}$, the unit of $L$ is $\rm{fm}$ and that of $r_v$ is $\rm{GeV^{-1}}$.}
\end{figure*}

The separation distance of a heavy-quark pair can  also be calculated from  Eq.~(\ref{distance1}). In Fig.~\ref{01ILE} (a), we can see that $L$ increases with an increase in $r_v$. There is an endpoint $r_v = 1.45 \rm{GeV^{-1}}$, where $\alpha$ tends to zero. Similarly by solving the force balance equation of~(\ref{force balance equation2}), we obtain numerical results for $\alpha$. It is found that $\alpha$ is a monotone function of $r_v$ and decreases with an increase in $r_v$. From Fig.~\ref{01ILE} (a), it is clear that the separate distance will increase with  increasing $r_v$. It should be noted that  $r_v$ has a maximum value beyond which the configuration will turn to the third case. The maximum separate distance and energy emerge at $r_v =1.45 \rm{GeV^{-1}}$. The corresponding energy  can also be evaluated, which is shown in Fig.~\ref{01ILE} (b). In this configuration, the energy is a linear function of $L$. Next, we turn to the third case and discuss it further.

\begin{center}
	\includegraphics[width=8cm]{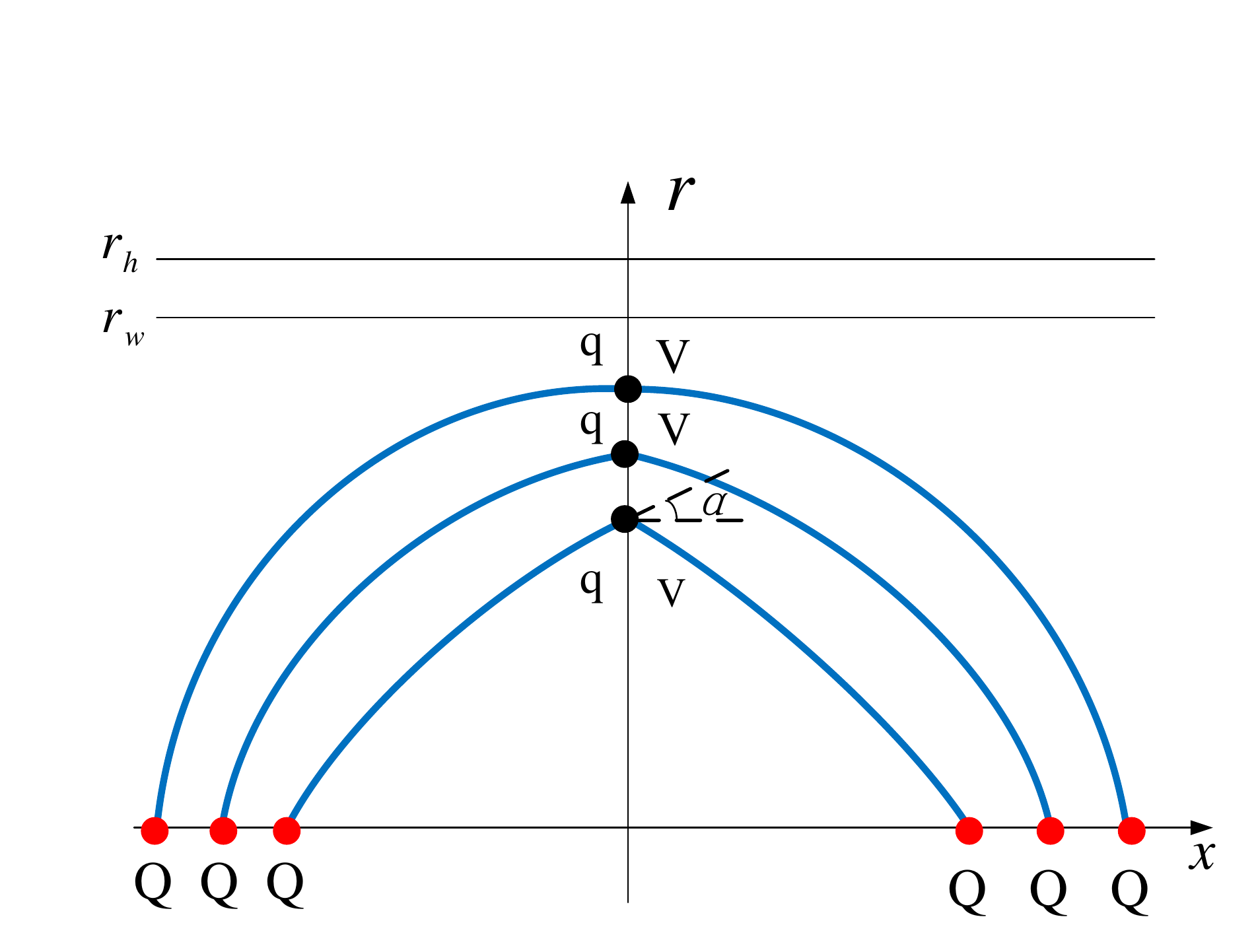}
	\figcaption{\label{01Sketch2} Schematic diagram of the string configuration with increasing $r_v$ for intermidiate $L$. $\alpha$ is always positive.}
\end{center}

\subsubsection{large $L$}
\begin{figure*}
	\centering
	\includegraphics[width=14cm]{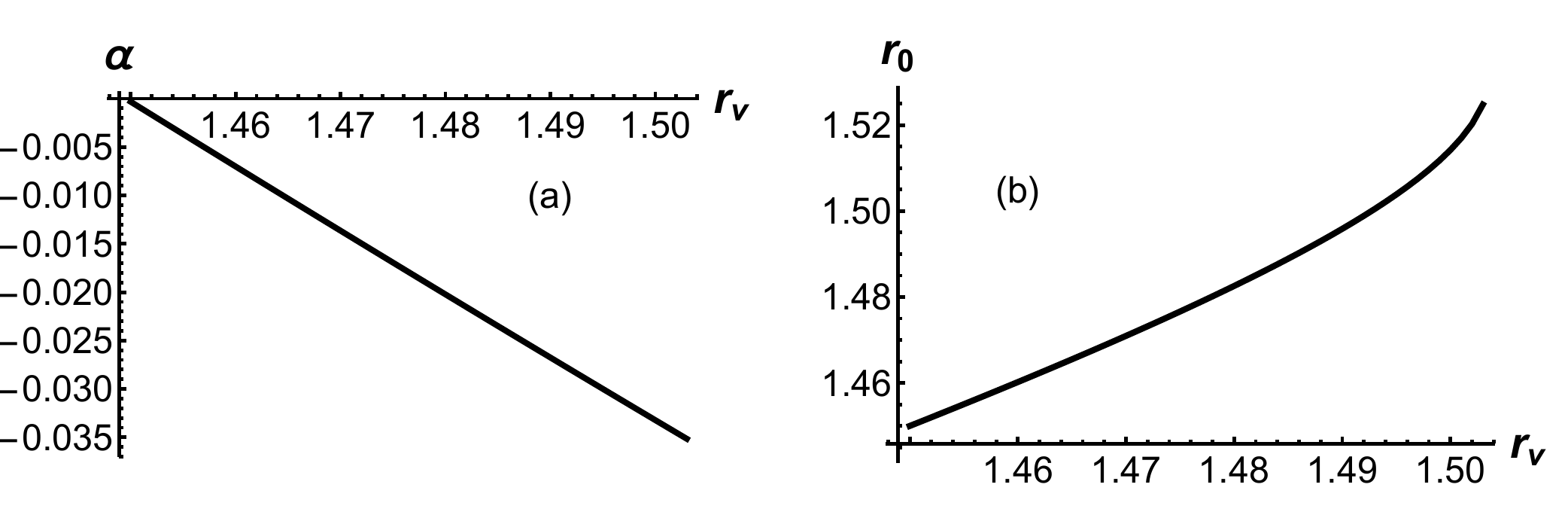}
	\caption{\label{01r0rv} (a)$\alpha$ as a function of $r_v$. $r_0$ as a function of $r_v$. $r_0$ and $r_v$ have the unit of $GeV^{-1}$.}
\end{figure*}

\begin{figure*}
	\centering
	\includegraphics[width=14cm]{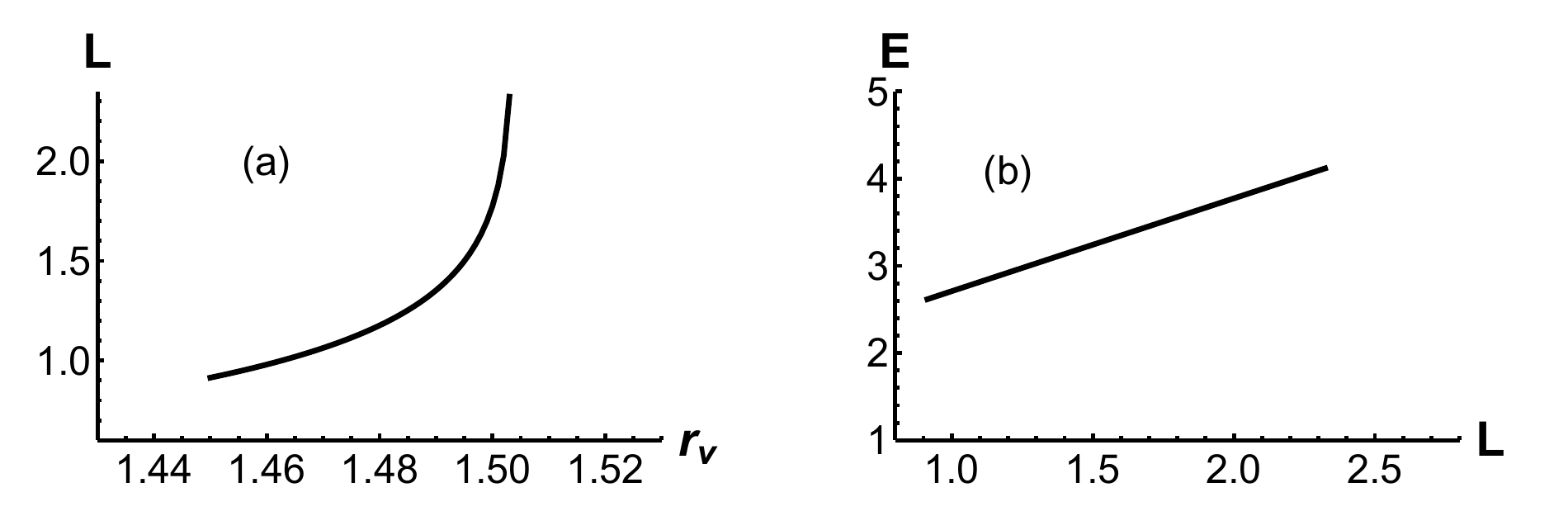}
	\caption{\label{01LLE} (a)Separation distance $L$ as a function of $r_v$. (b)The energy $E$ as a function of $L$. The unit of $E$ is $\rm{GeV}$, the unit of $L$ is $\rm{fm}$ and that of $r_v$ is $\rm{GeV^{-1}}$.}
\end{figure*}

Using Eq.~(\ref{force balance equation2}), we first numerically obtain the relationship between $\alpha$ and $r_v$ ,as shown in Fig.~\ref{01r0rv} (a). Unlike the previous case, the difference here is a negative $\alpha$ . Calculating the first integral from Eq.~(\ref{first2}), we find that $r_0$ is a function of $r_v$ in  Fig.~\ref{01r0rv} (b). As shown, the maximum  $r_v$ is $r_v \approx 1.48GeV^{-1}$ ,which corresponds to the position of the soft wall $r_w \approx r_0 \approx 1.53 GeV^{-1}$. We can also see that the separation distance tends to infinity when $r_0$ approaches $1.53 GeV^{-1}$ in Fig.~\ref{01LLE} (a).

Finally, the corresponding potential energy is shown in Fig.~\ref{01LLE} (b).Because the absolute value of $\alpha$ decreases with  increasing  $r_v$, we present a schematic diagram of third configuration in Fig.~\ref{01Sketch3}. At $r_0 \approx r_w$, $\alpha$ tends to zero ,and the separation distance becomes extremely large.

\begin{center}
	\includegraphics[width=8cm]{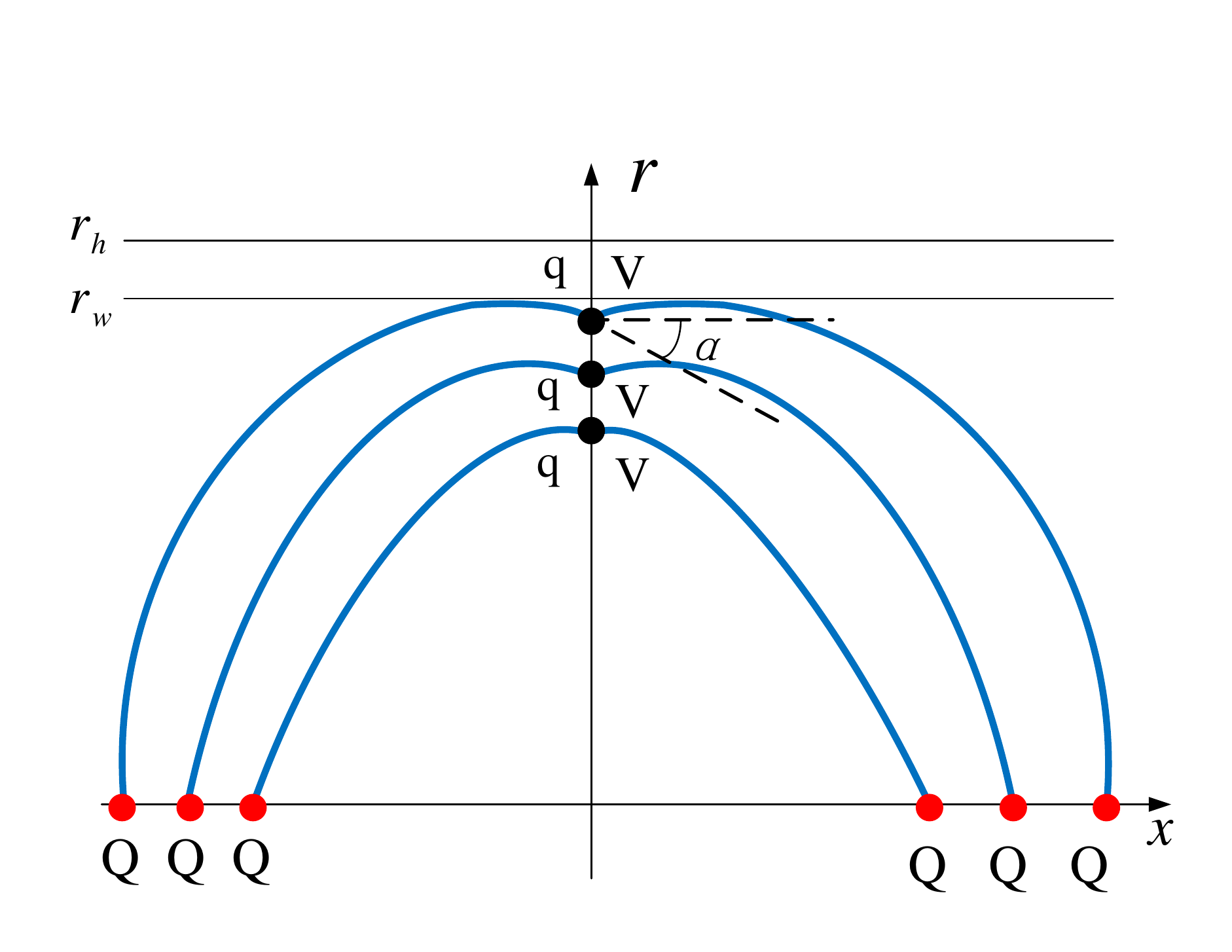}
	\figcaption{\label{01Sketch3} A schematic diagram of the string configuration with the increase of $r_v$ for large $L$. $\alpha$ is always negative.}
\end{center}

\subsubsection{Short summary}
If we combine all the configurations, we can present  the energy as a function of $L$ at all distances, as shown in Fig.~\ref{01EL}. Cleaqrly, the energy is smoothly increasing with separation distance for all configurations.

\begin{center}
	\includegraphics[width=8cm]{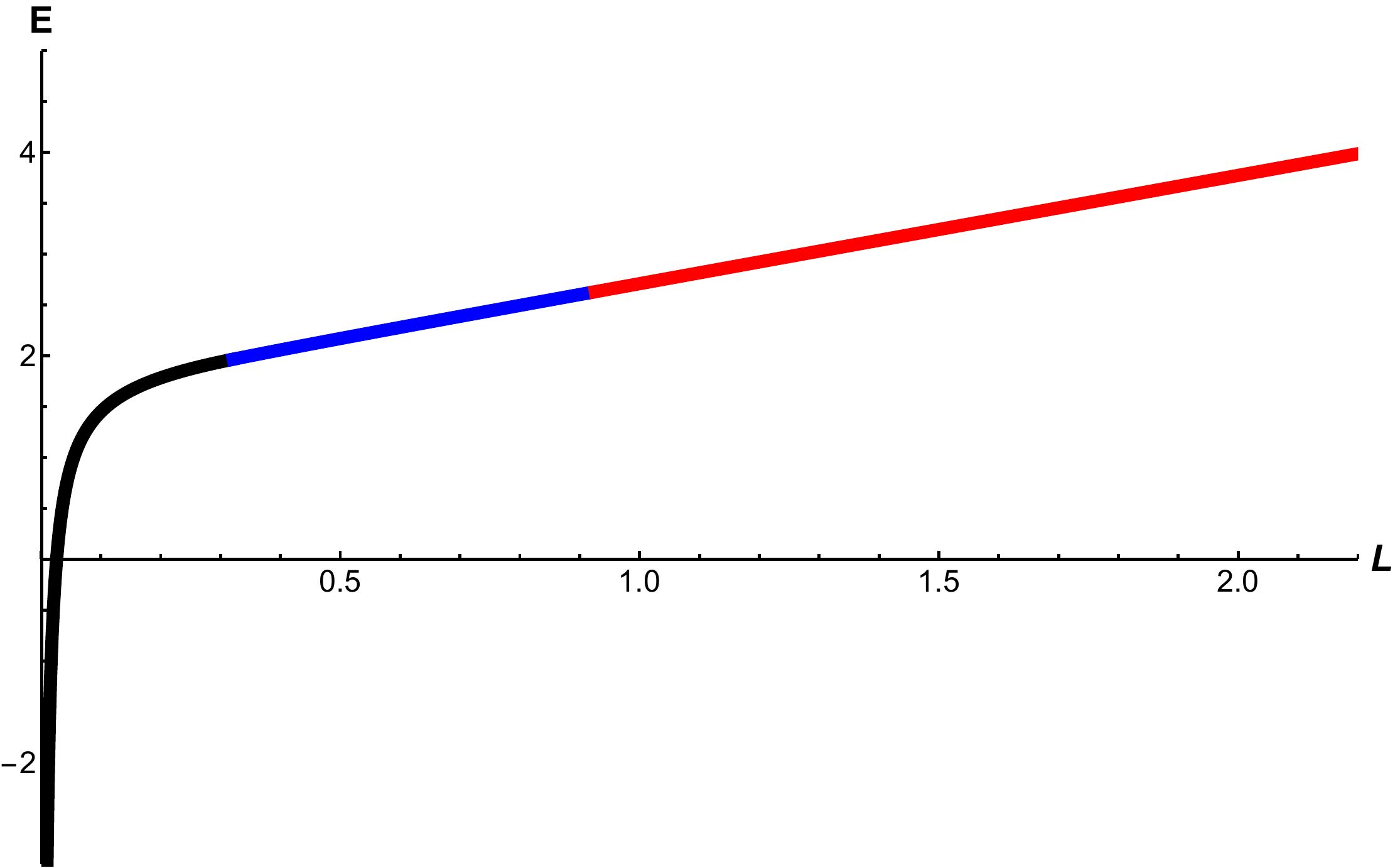}
	\figcaption{\label{01EL} Energy $E$ as a function of $L$ at all separation distances. The black line represents small distances, the blue line represents intermediate distances, and the red line represents large distances. The unit of $E$ is $\rm{GeV}$,and the unit of $L$ is $\rm{fm}$.}
\end{center}

\subsection{$T=0.148GeV$}
At temperature $T=0.148GeV$,  the system is in the deconfined phase. In this phase, the soft wall disappears, and the QQq will melt at a sufficient distance.

\subsubsection{small $L$}
\begin{figure*}
	\centering
	\includegraphics[width=14cm]{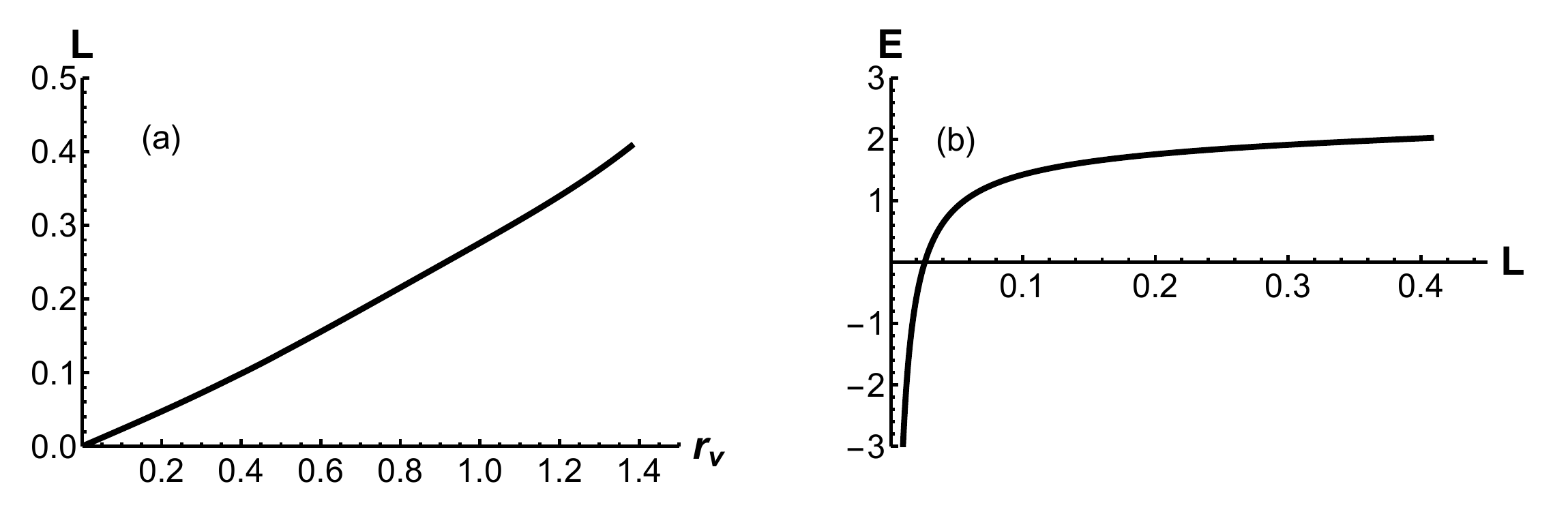}
	\caption{\label{0148SLE} (a)Separate distance $L$ as a function of $r_v$. (b)The energy $E$ as a function of $L$.  The unit of $E$ is $\rm{GeV}$, $L$ is $\rm{fm}$ and $r_v$ is $\rm{GeV^{-1}}$.}
\end{figure*}

First, we can  determine the position of the light quark from~(\ref{lightquarkforce}). At $T=0.148GeV$, we find $r_q = 1.38 GeV^{-1}$ or $r_q = 1.54 GeV^{-1}$. In this paper, we focus on the ground state and only consider $r_q = 1.38 GeV^{-1}$ at $T=0.148GeV$. The angle $\alpha$ can be calculated from Eq.~(\ref{force balance equation}). The energy and separation distance are calculated using the same procedure as before, and the results are shown in Fig.~\ref{0148SLE}. $L$ is still an increasing function of $r_v$ ,and $E$ is a Cornell-like potential.

\subsubsection{Intermediate $L$}
\begin{figure*}
	\centering
	\includegraphics[width=14cm]{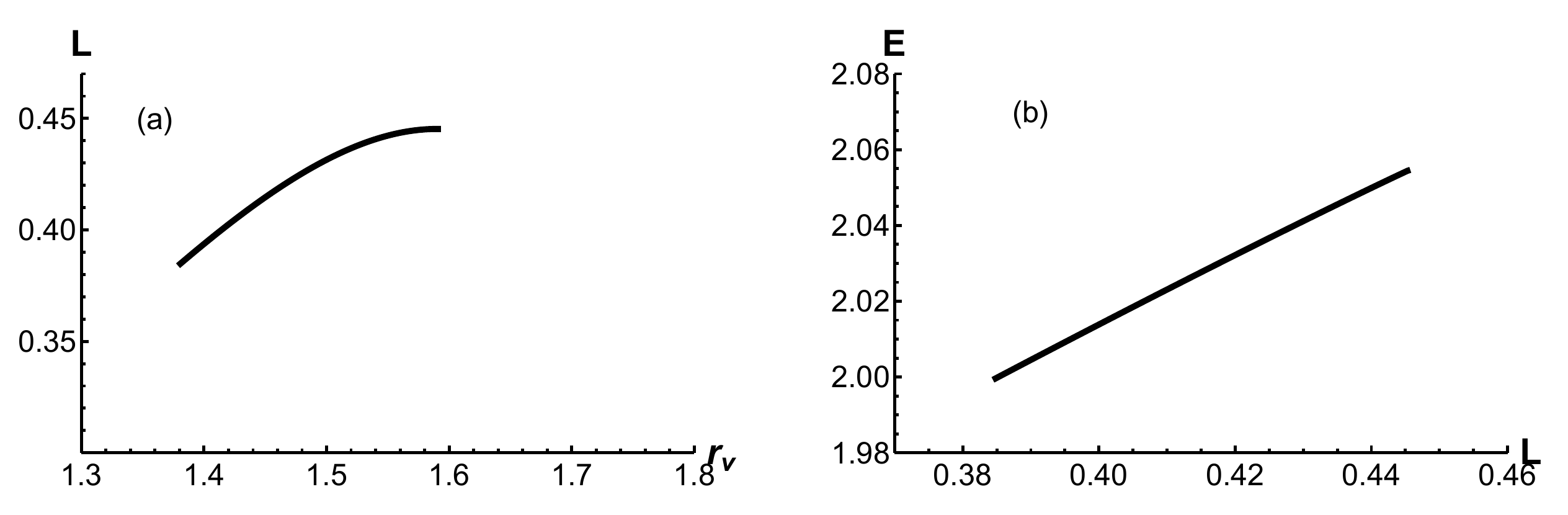}
	\caption{\label{0148ILE} (a)Separation distance $L$ as a function of $r_v$. (b)The energy $E$ as a function of $L$.  The unit of $E$ is $\rm{GeV}$, the unit of $L$ is $\rm{fm}$ and that of $r_v$ is $\rm{GeV^{-1}}$.}
\end{figure*}

In this configuration, we calculate the energy and separation distance in Fig.~\ref{0148ILE}. With an increasing in $r_v$, the separation distance increases. There is a maximum value $L_{max} = 0.445 \rm{fm}$ beyond which the configuration can not exist and quarks become free. Thus, we find that the third configuration can't exist at $T= 0.145 GeV$.

\subsubsection{Short summary}
Only the first and second configurations can exist at $T = 0.148GeV$. We present the energy as a function of  separation distance from 0 to $L_{max}$ in Fig.~\ref{0145EL}. The potential is smoothly increasing from small distances to intermediate distances. The potential ends at $L_{max} = 0.445 \rm{fm}$, marked by the red dot in the figure. At large distances,  QQq melts and becomes free quarks.
\begin{center}
	\includegraphics[width=8cm]{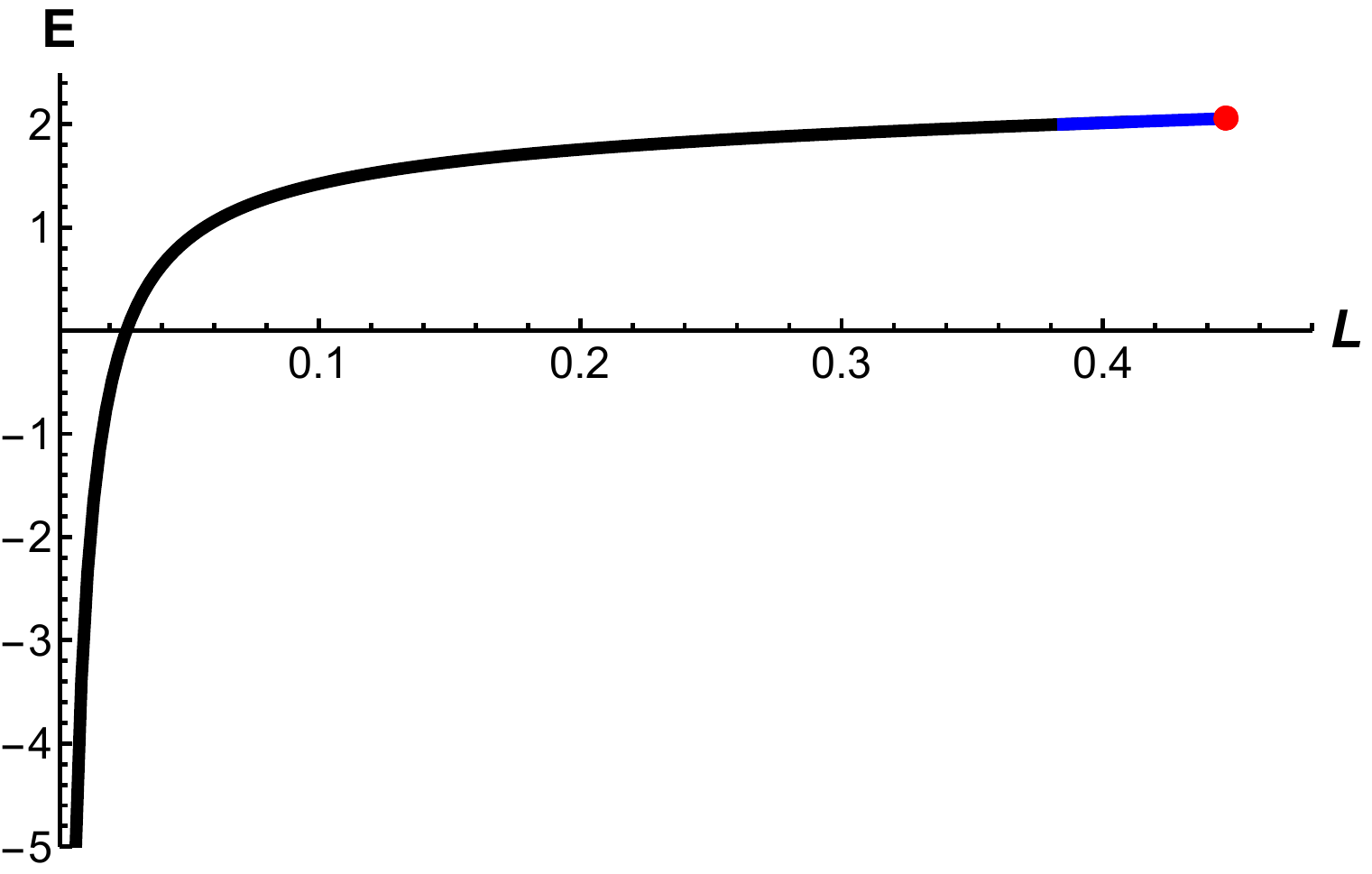}
	\figcaption{\label{0145EL} Energy $E$ as a function of $L$ at all distances. The black line represents $E_{QQq}$ at small distances,the blue line represents $E_{QQq}$ at intermediate distances,and the red line represents $E_{QQq}$ at large distances. The unit of $E$ is $\rm{GeV}$, and the unit of $L$ is $\rm{fm}$.}
\end{center}

\subsection{$T=0.15GeV$}
\begin{figure*}
	\centering
	\includegraphics[width=14cm]{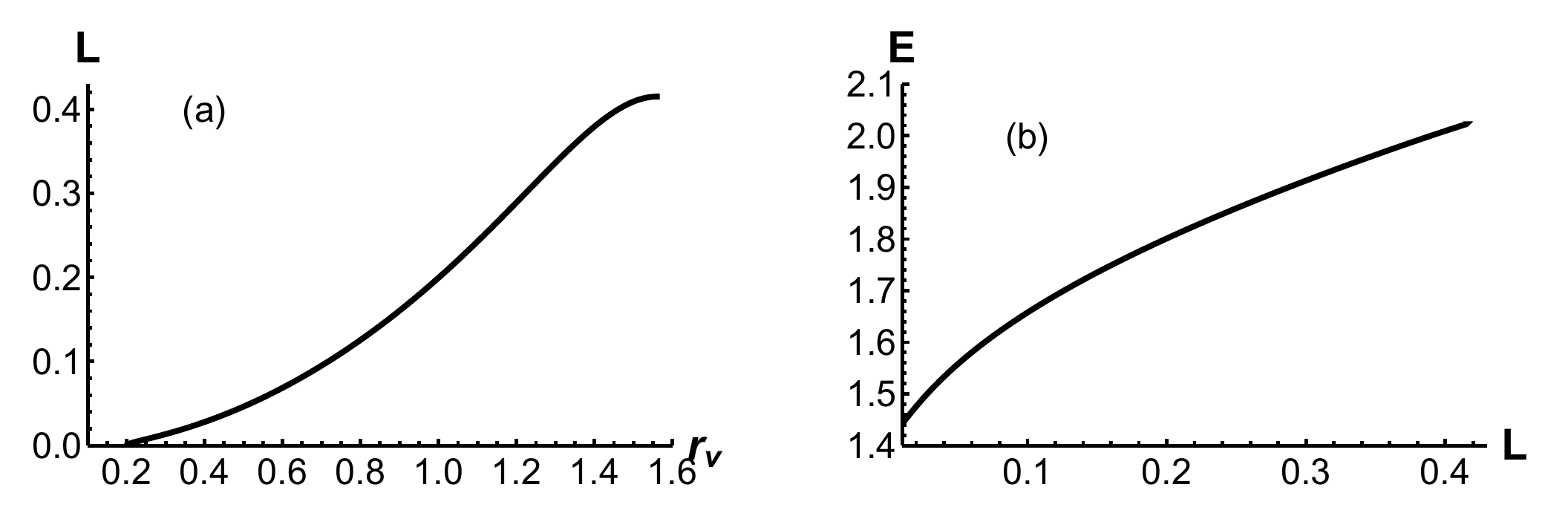}
	\caption{\label{015ILE}(a)Separation distance $L$ as a function of $r_v$. (b)The energy $E$ as a function of $L$. The unit of $E$ is $\rm{GeV}$, the unit of$L$ is $\rm{fm}$ and that of $r_v$ is $\rm{GeV^{-1}}$.}
\end{figure*}

First, we check the force balance equation of the light quark for the first configuration and find there is no solution for any $r_q$ at $T=0.15GeV$, as shown in Fig.~\ref{force150}. Thus, the first configuration can not  exist at this temperature. From Eq.~(\ref{force balance equation2}), we can find the relationship between $\alpha$ and $r_v$. There is also a maximum value $r_{max}$ at $T=0.15GeV$, beyond which QQq melts as shown in Fig.~\ref{020sketch}. The separation distance and energy of the string are shown in Fig.~\ref{015ILE}.

\begin{center}
	\includegraphics[width=8cm]{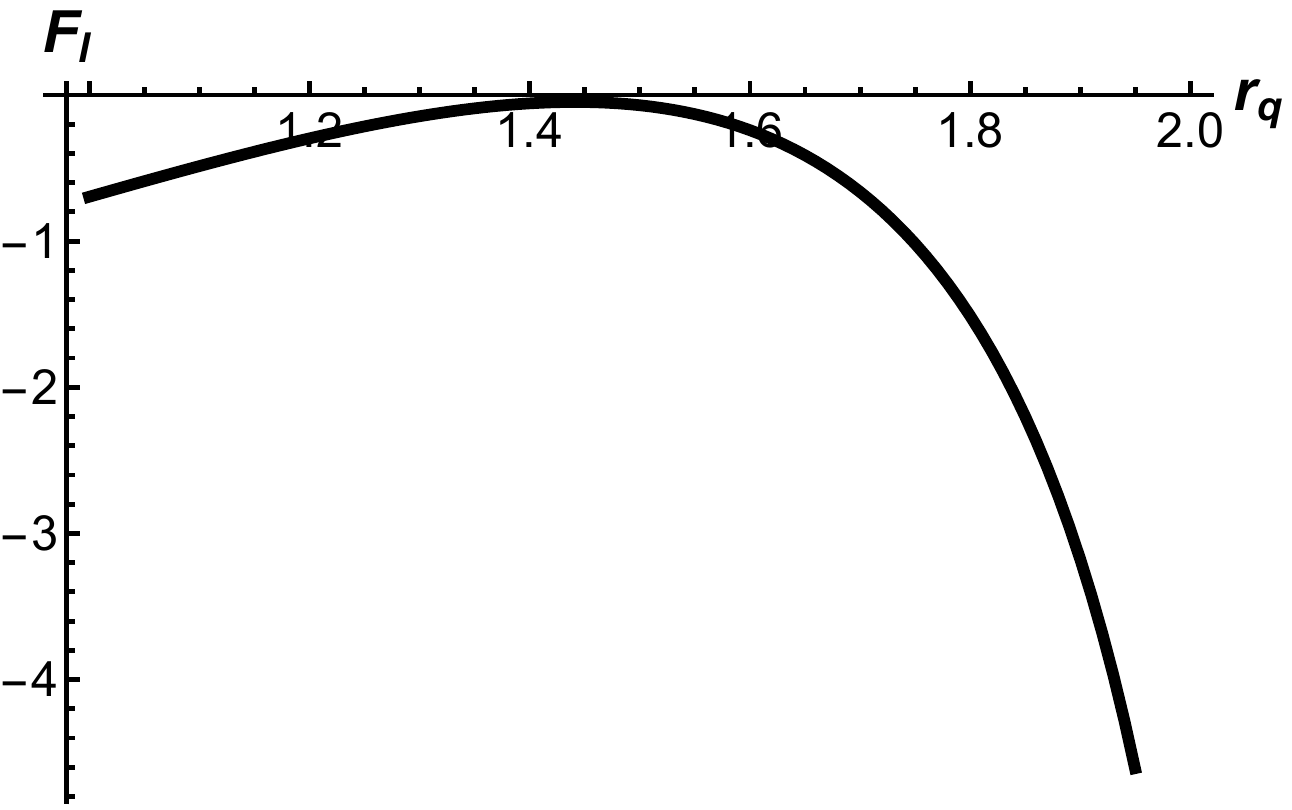}
	\figcaption{\label{force150}  Force balance equation of the light quark as a function of $r_q$. The unit of $r_q$ is $GeV^{-1}$}
\end{center}

\begin{center}
	\includegraphics[width=8cm]{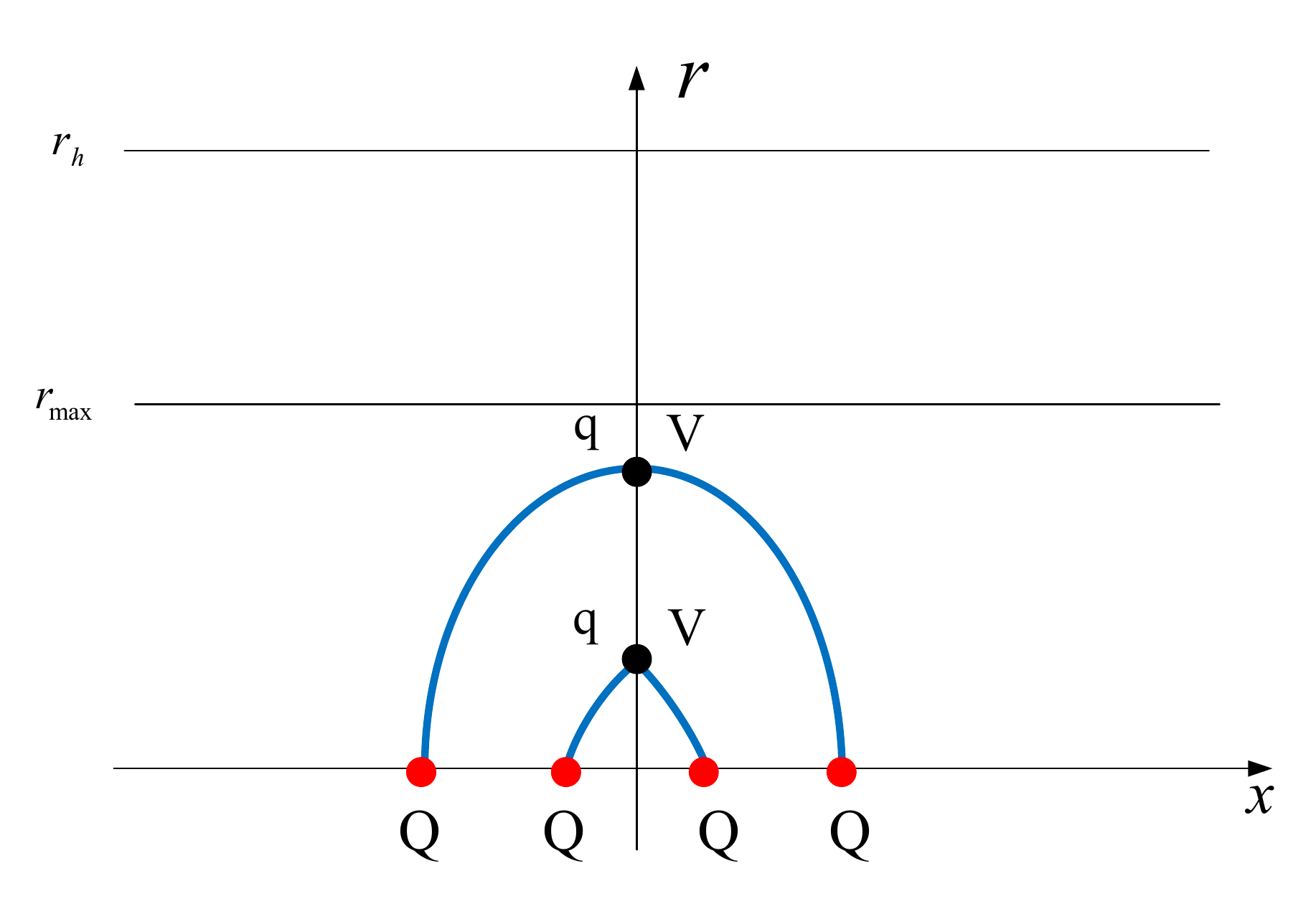}
	\figcaption{\label{020sketch} Schematic diagram of the string configuration with  increasing  $r_v$.}
\end{center}

\section{String breaking in the confined phase}\label{breaking}
In the confined phase, the quarks are confined in the hadrons. Can QQq exist at extremely large distances? The answer is no. The light quarks and anti-quark will be excited from vacuum at large distances. We call this string breaking and consider the following decay mode

\begin{equation}
Q Q q \rightarrow Q q q+Q \bar{q}.
\end{equation}
Qqq consists of three fundamental strings: a vertex and two light quarks. Thus, the total action of Qqq is $S=\sum_{i=1}^{3} S_{\mathrm{NG}}^{(i)}+S_{\mathrm{vert}}+2S_{\mathrm{q}}$.  $Q\bar{q}$  consists of a fundamental string and a light quark. Thus, the total action of Qqq is $S= S_{\mathrm{NG}}+S_{\mathrm{q}}$. The total actions of $Qqq$ and $Q\bar{q}$ are
\begin{equation}
\begin{aligned}
S_{Qqq} &= \mathbf{g} \Big(2 \int_{r_v}^{r_q} \frac{e^{s r^2}}{r^2} dr +  \int_{0}^{r_v} \frac{e^{s r^2}}{r^2}+ 3k  \frac{e^{-2 s r_v^2} \sqrt{f(r_v)}}{r_v}\\&+2 n  \frac{e^{\frac{1}{2}s r_q^2}}{r_q} \sqrt{f(r_q)}\Big), \\
S_{Q\bar{q}} &= \mathbf{g} \int_{0}^{r_q}\frac{{e^{s r^2}}}{r^2} +  n \mathbf{g} \frac{e^{\frac{1}{2}s r_q^2}}{r_q} \sqrt{f(r_q)} .
\end{aligned}
\end{equation}
Varying the action with respect to $r_q$ gives  Eq.~(\ref{lightquarkforce}), and varying the action with respect to $r_v$ gives

\begin{equation}
\begin{aligned}
1 + 3k e^{-3 s r_v^3} \sqrt{f(r_v)}&+ 12 k s e^{-3 s r_v^3} r_v^2\sqrt{f(r_v)}\\&-
\frac{3 k e^{-3 s r_v^2} r_v f'(r_v)}{2\sqrt{f(r_v)}} = 0.
\end{aligned}
\end{equation}

We can obtain $r_v=0.410 GeV^{-1}$ or $r_v=0.453 GeV^{-1}$. Because the difference in energy is extremely small for the two states, we take $r_v=0.410 GeV^{-1}$ for simplicity. The configuration for $Q q q+Q \bar{q}$ is shown in Fig.~\ref{sketch}. The renormalized total energy is
\begin{equation}
\begin{aligned}
E_{Qqq}+E_{Q\bar{q}} &= \mathbf{g} \Big(2 \int_{r_v}^{r_q} \frac{e^{s r^2}}{r^2} dr + \int_{0}^{r_q}(\frac{{e^{s r^2}}}{r^2} - \frac{1}{r^2}) - \frac{1}{r_q} + \\& \int_{0}^{r_v} (\frac{e^{s r^2}}{r^2}-\frac{1}{r^2})- \frac{1}{r_v}+ 3k  \frac{e^{-2 s r_v^2} \sqrt{f(r_v)}}{r_v}\\&+ 3 n  \frac{e^{\frac{1}{2}s r_q^2}}{r_q} \sqrt{f(r_q)}\Big) + 2c.
\end{aligned}
\end{equation}
For fixed $r_q=1.146 GeV^{-1}$ and $r_v=0.410 GeV^{-1}$, we have $E_{Qqq}+E_{Q\bar{q}} = 3.006 GeV$. Fig.\ref{sketch2} is a schematic diagram of string breaking.To determine the string breaking distance, we plot the energy $E_{QQq}$ and $E_{Q\bar{q}}+ E_{Qqq}$ as a function of $r_v$ at $T = 0.1GeV$ in Fig.~\ref{totalpotential}. The cross point shown in the figure enables us to determine $L_{QQq}$ at fixed temperature. We find the distance of string breaking to be  $L_{QQq} = 1.27 \rm{fm}$.
\begin{center}
	\includegraphics[width=8cm]{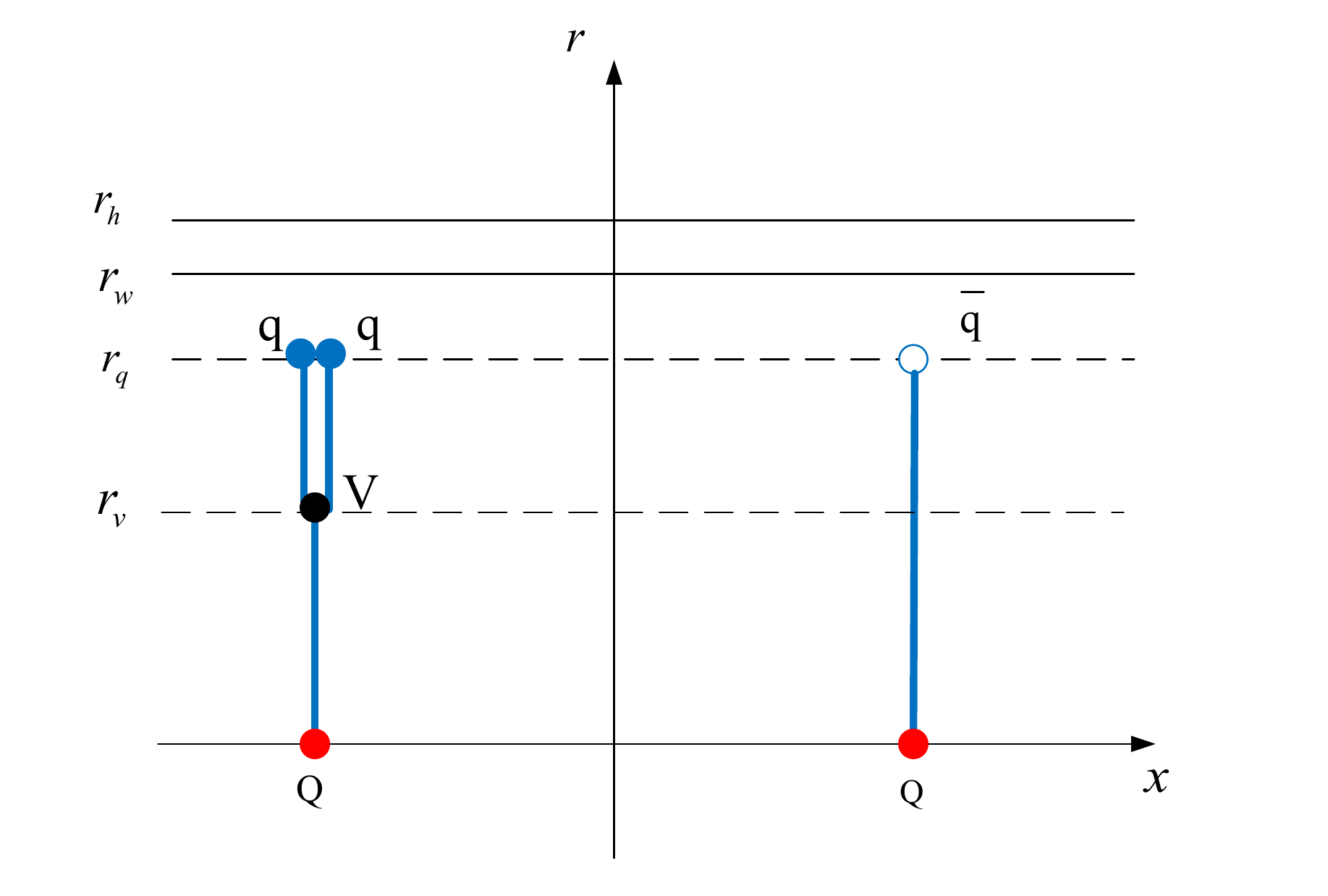}
	\figcaption{\label{sketch} A schematic diagram of $Q q q+Q \bar{q}$.}
\end{center}

\begin{center}
	\includegraphics[width=8cm]{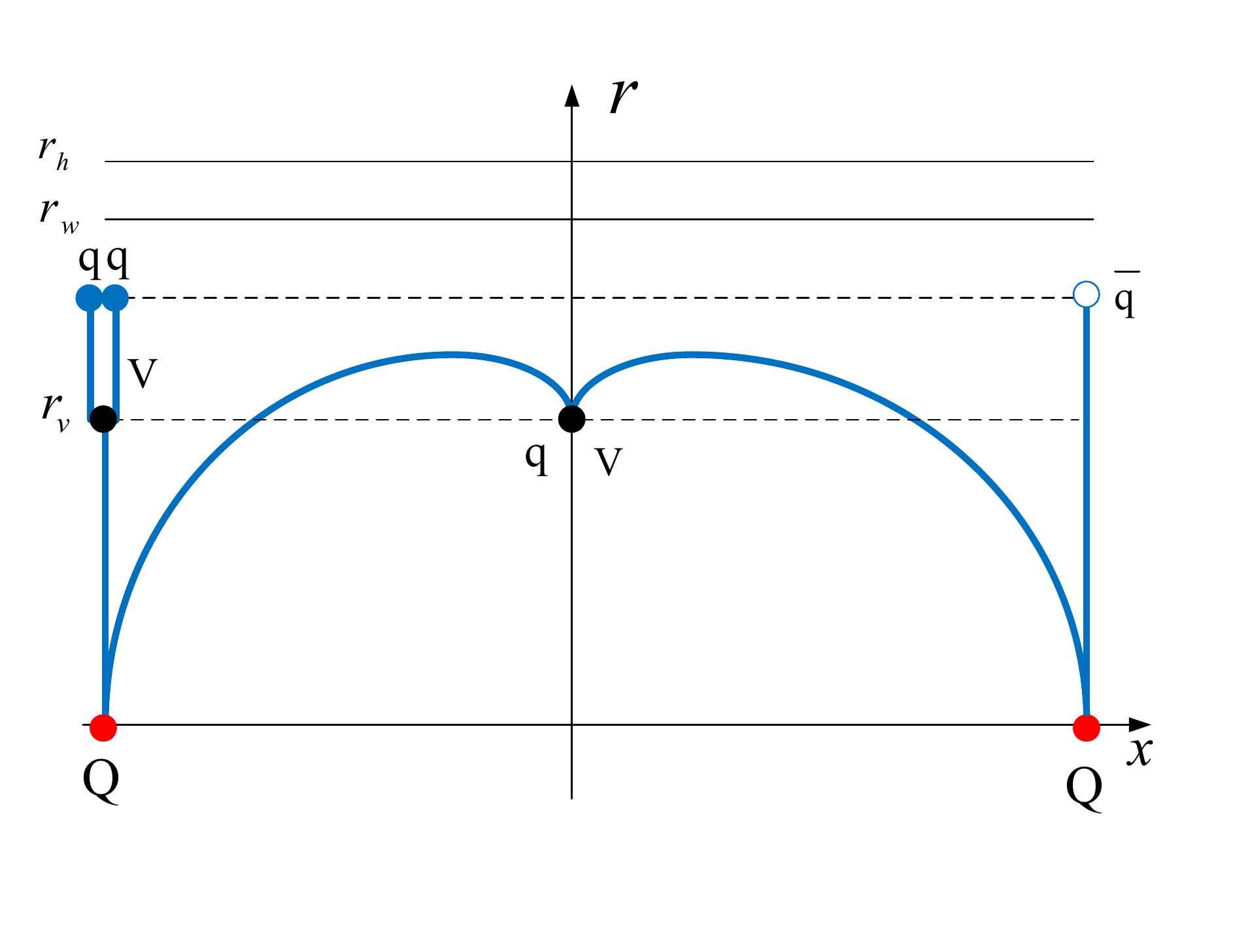}
	\figcaption{\label{sketch2} Schematic diagram of  string breaking from the third configuration of QQq to $\rm Q q q+Q \bar{q}$.}
\end{center}

\begin{center}
	\includegraphics[width=8cm]{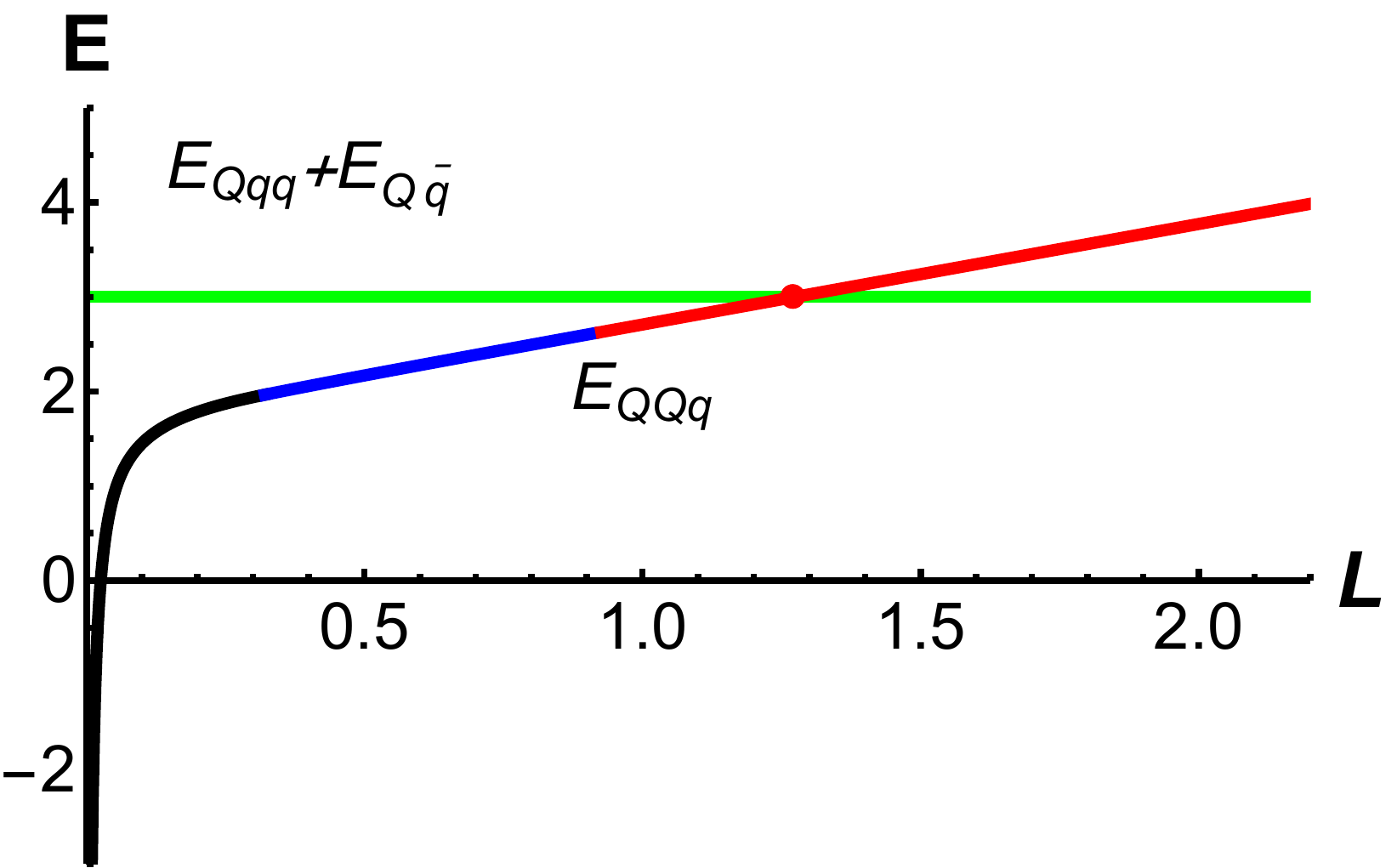}
	\figcaption{\label{totalpotential} Green line is the energy $E_{\rm{Qqq}}+E_{\rm{Q\bar{q}}}$. The black line represents $E_{QQq}$ at small distances,the blue line represents $E_{QQq}$ at intermediate distances,and the red line represents $E_{QQq}$ at large distances. The unit of $E$ is GeV and the unit of $r_v$ is $\rm GeV^{-1}$.}
\end{center}

\section{COMPARING WITH $\rm{Q\bar{Q}}$}\label{compare}
The energy of $\rm{Q\bar{Q}}$ has been extensively studied in many holographic models. In this section, we focus on comparing the energy of $\rm{Q\bar{Q}}$ with QQq in the confined and deconfined phases. First, we only show the results of the separation distance and energy of $\rm{Q\bar{Q}}$

\begin{equation}
\begin{aligned}
L_{Q\bar{Q}} = 2 \int_{0}^{r_0} (\frac{g_2(r)}{g_1(r)}(\frac{g_2(r)}{g_2(r_0)}-1))^{-\frac{1}{2}},
\end{aligned}
\end{equation}

\begin{equation}
\begin{aligned}
E_{Q\bar{Q}} =2 \mathbf{g} (\int_{0}^{z_0} \sqrt{\frac{g_2(r) g_1(r)}{g_2(r)-g_2(r_0)}}-\frac{1}{r^2}) - \frac{2 \mathbf{g}}{r_0} + 2c,
\end{aligned}
\end{equation}
where $g_1(r) = \frac{e^{2 s r^2}}{r^4}, g_2(r) = \frac{e^{2 s r^2}}{r^4} f(r)$, and $z_0$ is the turning point of the U-shape string.A detailed calculation can be found in our previous paper\cite{Chen:2017lsf,Zhou:2020ssi,Zhou:2021sdy}. We consider the decay mode $\rm{Q \bar{Q}\rightarrow Q\bar{q} + \bar{Q}q}.$ It is clear that

\begin{equation}
\begin{aligned}
E_{Q\bar{q}} = \mathbf{g} (\int_{0}^{r_q} \frac{1}{r^2}(e^{s r^2} - 1)) -\frac{\mathbf{g}}{r_q} + \mathbf{g} n \frac{e^{\frac{s}{2}r_q^2}}{rq}\sqrt{f(r_q)} + c.
\end{aligned}
\end{equation}

Thus, we can calculate $E_{\rm{Q\bar{q}}}+E_{\rm{\bar{Q}q}} = 2.39 GeV$ at $T = 0.1GeV$. In the confined phase, such as when $T = 0.1GeV$, we present the energy of $E_{\rm{Q\bar{q}}}+E_{\rm{\bar{Q}q}}$ and $E_{\rm{Q\bar{Q}}}$ in Fig.~\ref{QQpotential}.    Trough a  comparison with Fig.~\ref{totalpotential}, we find the distance of string breaking $L=1.25 \rm{fm}$ is close to that of QQq.

\begin{center}
	\includegraphics[width=8cm]{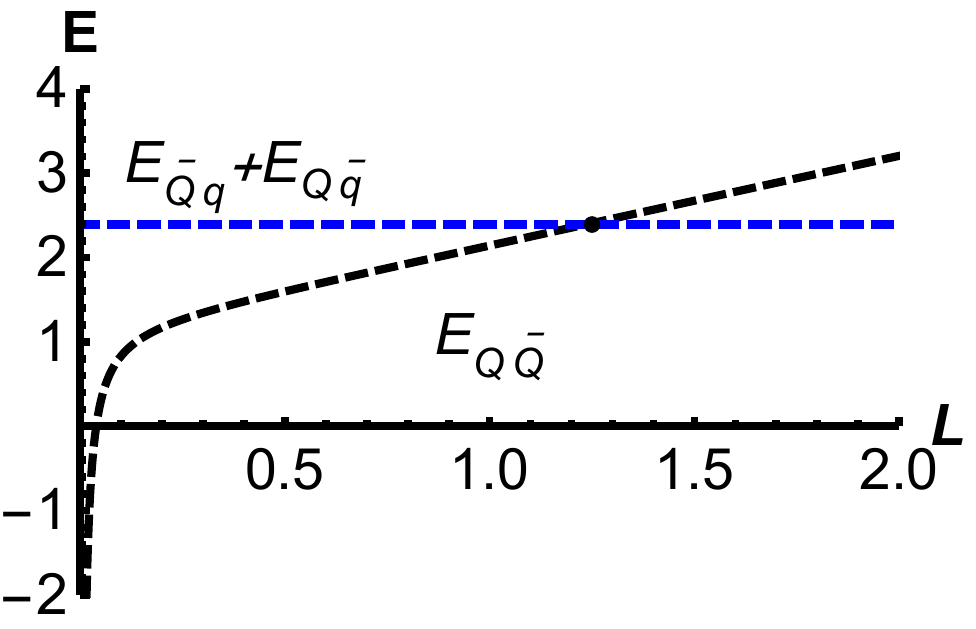}
	\figcaption{\label{QQpotential}  Blue dashed line is the energy of $E_{Q\bar{q}}+E_{\bar{Q}q}$ ,and the black dashed line is the energy of $E_{Q\bar{Q}}$. The temperature is $T = 0.1\rm{GeV}$. The unit of $E$ is GeV ,and the unit of $L$ is $\rm fm$.}
\end{center}

In the deconfinement phase, we compare the energies of $E_{\rm{Q\bar{Q}}}$ and $E_{\rm{QQq}}$ at $T = 0.148GeV$. Fig.~\ref{QQpotential2} shows that the screening distance of $E_{\rm{QQq}}$($L = 0.45\rm{fm}$) is significantly smaller than that of $E_{Q\bar{Q}}$($L = 0.94\rm{fm}$) at the same temperature $T = 1.48 \rm{GeV}$.This indicates that $\rm{Q\bar{Q}}$ is more stable than QQq in the deconfined phase.

\begin{center}
	\includegraphics[width=8cm]{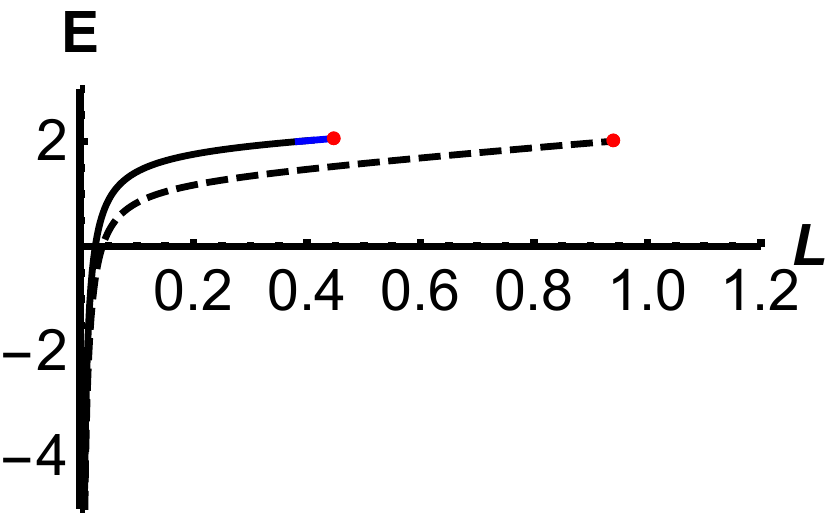}
	\figcaption{\label{QQpotential2}  Solid line is the energy of $E_{\rm{QQq}}$, and the dashed line is the energy of $E_{\rm{Q\bar{Q}}}$. The temperature is $T = 0.148\rm{GeV}$. The unit of $E$ is GeV, and the unit of $L$ is $\rm fm$.}
\end{center}

\end{multicols}
\begin{multicols}{2}

\section{SUMMARY AND CONCLUSION}\label{Summary}
In this paper, we mainlyfocused on QQq melting and string breaking at finite temperature through a five-dimensional effective string model.
For the confined phase, string and light quarks can not exceed the soft wall. Quarks are permanently confined in the hadrons.
However,  string breaking of QQq occurs at sufficiently large distances. We considered the decay mode $\rm{Q Q q \rightarrow Q q q+Q \bar{q}}$  and found the distance of string breaking to be  $L = 1.27 \rm{fm}$.
Other decay modes may be considered in the future studies. At a high temperature, the system is in the deconfined phase which implies that the soft wall disappears and QQq  melts at a certain distance.
For example, QQq  melts at $L = 0.45\rm{fm}$ for $T = 0.148GeV$. In contrast, we found that  $\rm{Q\bar{Q}}$  melts at $L = 0.94\rm{fm}$ for $T = 0.148GeV$, which indicates that $\rm{Q\bar{Q}}$ is more stable than QQq at high temperatures.
Finally, we hope that studying  effective string model will provide more observable quantities for future experiments .

\section{Acknowledgments}

\acknowledgments{This work is supported by the NSFC under Grants No. 12175100, No. 11975132 and the Research Foundation of Education Bureau of Hunan Province, China (Grant No. 21B0402, No. 21A0280 and No. 20C1594).}

\end{multicols}

\vspace{10mm}

\vspace{-1mm}
\centerline{\rule{80mm}{0.1pt}}
\vspace{2mm}

\begin{multicols}{2}

\end{multicols}

\clearpage

\end{CJK*}

\begin{thebibliography}{90}

\vspace{3mm}
%1\cite{M. Bashkanov:2009}

%\cite{Maldacena:1998im}
\bibitem{Maldacena:1998im}
J.~M.~Maldacena,
%``Wilson loops in large N field theories,''
Phys. Rev. Lett. \textbf{80}, 4859-4862 (1998)
doi:10.1103/PhysRevLett.80.4859
[arXiv:hep-th/9803002 [hep-th]].
%1771 citations counted in INSPIRE as of 29 Aug 2021

%\cite{Rey:1998bq}
\bibitem{Rey:1998bq}
S.~J.~Rey, S.~Theisen and J.~T.~Yee,
%``Wilson-Polyakov loop at finite temperature in large N gauge theory and anti-de Sitter supergravity,''
Nucl. Phys. B \textbf{527}, 171-186 (1998)
doi:10.1016/S0550-3213(98)00471-4
[arXiv:hep-th/9803135 [hep-th]].
%445 citations counted in INSPIRE as of 29 Aug 2021

%\cite{Brandhuber:1998bs}
\bibitem{Brandhuber:1998bs}
A.~Brandhuber, N.~Itzhaki, J.~Sonnenschein and S.~Yankielowicz,
%``Wilson loops in the large N limit at finite temperature,''
Phys. Lett. B \textbf{434}, 36-40 (1998)
doi:10.1016/S0370-2693(98)00730-8
[arXiv:hep-th/9803137 [hep-th]].
%324 citations counted in INSPIRE as of 29 Aug 2021

%\cite{Andreev:2006ct}
\bibitem{Andreev:2006ct}
O.~Andreev and V.~I.~Zakharov,
%``Heavy-quark potentials and AdS/QCD,''
Phys. Rev. D \textbf{74}, 025023 (2006)
doi:10.1103/PhysRevD.74.025023
[arXiv:hep-ph/0604204 [hep-ph]].
%238 citations counted in INSPIRE as of 29 Aug 2021


%\cite{Andreev:2006eh}
\bibitem{Andreev:2006eh}
O.~Andreev and V.~I.~Zakharov,
%``The Spatial String Tension, Thermal Phase Transition, and AdS/QCD,''
Phys. Lett. B \textbf{645}, 437-441 (2007)
doi:10.1016/j.physletb.2007.01.002
[arXiv:hep-ph/0607026 [hep-ph]].
%111 citations counted in INSPIRE as of 29 Aug 2021

%\cite{Andreev:2006nw}
\bibitem{Andreev:2006nw}
O.~Andreev and V.~I.~Zakharov,
%``On Heavy-Quark Free Energies, Entropies, Polyakov Loop, and AdS/QCD,''
JHEP \textbf{04}, 100 (2007)
doi:10.1088/1126-6708/2007/04/100
[arXiv:hep-ph/0611304 [hep-ph]].
%60 citations counted in INSPIRE as of 29 Aug 2021


%\cite{He:2010bx}
\bibitem{He:2010bx}
S.~He, M.~Huang and Q.~s.~Yan,
%``Heavy quark potential and QCD beta function from a deformed $AdS_5$ model,''
Prog. Theor. Phys. Suppl. \textbf{186}, 504-509 (2010)
doi:10.1143/PTPS.186.504
[arXiv:1007.0088 [hep-ph]].
%6 citations counted in INSPIRE as of 29 Aug 2021

%\cite{Colangelo:2010pe}
\bibitem{Colangelo:2010pe}
P.~Colangelo, F.~Giannuzzi and S.~Nicotri,
%``Holography, Heavy-Quark Free Energy, and the QCD Phase Diagram,''
Phys. Rev. D \textbf{83}, 035015 (2011)
doi:10.1103/PhysRevD.83.035015
[arXiv:1008.3116 [hep-ph]].
%67 citations counted in INSPIRE as of 29 Aug 2021


%\cite{DeWolfe:2010he}
\bibitem{DeWolfe:2010he}
O.~DeWolfe, S.~S.~Gubser and C.~Rosen,
%``A holographic critical point,''
Phys. Rev. D \textbf{83}, 086005 (2011)
doi:10.1103/PhysRevD.83.086005
[arXiv:1012.1864 [hep-th]].
%120 citations counted in INSPIRE as of 29 Aug 2021


%\cite{Li:2011hp}
\bibitem{Li:2011hp}
D.~Li, S.~He, M.~Huang and Q.~S.~Yan,
%``Thermodynamics of deformed AdS$_5$ model with a positive/negative quadratic correction in graviton-dilaton system,''
JHEP \textbf{09}, 041 (2011)
doi:10.1007/JHEP09(2011)041
[arXiv:1103.5389 [hep-th]].
%77 citations counted in INSPIRE as of 29 Aug 2021


%\cite{Fadafan:2011gm}
\bibitem{Fadafan:2011gm}
K.~B.~Fadafan,
%``Heavy quarks in the presence of higher derivative corrections from AdS/CFT,''
Eur. Phys. J. C \textbf{71}, 1799 (2011)
doi:10.1140/epjc/s10052-011-1799-7
[arXiv:1102.2289 [hep-th]].
%26 citations counted in INSPIRE as of 29 Aug 2021

%\cite{Fadafan:2012qy}
\bibitem{Fadafan:2012qy}
K.~B.~Fadafan and E.~Azimfard,
%``On meson melting in the quark medium,''
Nucl. Phys. B \textbf{863}, 347-360 (2012)
doi:10.1016/j.nuclphysb.2012.05.022
[arXiv:1203.3942 [hep-th]].
%17 citations counted in INSPIRE as of 29 Aug 2021

%\cite{Cai:2012xh}
\bibitem{Cai:2012xh}
R.~G.~Cai, S.~He and D.~Li,
%``A hQCD model and its phase diagram in Einstein-Maxwell-Dilaton system,''
JHEP \textbf{03}, 033 (2012)
doi:10.1007/JHEP03(2012)033
[arXiv:1201.0820 [hep-th]].
%63 citations counted in INSPIRE as of 29 Aug 2021


%\cite{Li:2012ay}
\bibitem{Li:2012ay}
D.~Li, M.~Huang and Q.~S.~Yan,
%``A dynamical soft-wall holographic QCD model for chiral symmetry breaking and linear confinement,''
Eur. Phys. J. C \textbf{73}, 2615 (2013)
doi:10.1140/epjc/s10052-013-2615-3
[arXiv:1206.2824 [hep-th]].
%53 citations counted in INSPIRE as of 29 Aug 2021


%\cite{Fang:2015ytf}
\bibitem{Fang:2015ytf}
Z.~Fang, S.~He and D.~Li,
%``Chiral and Deconfining Phase Transitions from Holographic QCD Study,''
Nucl. Phys. B \textbf{907}, 187-207 (2016)
doi:10.1016/j.nuclphysb.2016.04.003
[arXiv:1512.04062 [hep-ph]].
%28 citations counted in INSPIRE as of 29 Aug 2021


%\cite{Yang:2015aia}
\bibitem{Yang:2015aia}
Y.~Yang and P.~H.~Yuan,
%``Confinement-deconfinement phase transition for heavy quarks in a soft wall holographic QCD model,''
JHEP \textbf{12}, 161 (2015)
doi:10.1007/JHEP12(2015)161
[arXiv:1506.05930 [hep-th]].
%44 citations counted in INSPIRE as of 29 Aug 2021


%\cite{Zhang:2015faa}
\bibitem{Zhang:2015faa}
Z.~q.~Zhang, D.~f.~Hou and G.~Chen,
%``Heavy quark potential from deformed $AdS_5$ models,''
Nucl. Phys. A \textbf{960}, 1-10 (2017)
doi:10.1016/j.nuclphysa.2017.01.007
[arXiv:1507.07263 [hep-ph]].
%3 citations counted in INSPIRE as of 29 Aug 2021


%\cite{Ewerz:2016zsx}
\bibitem{Ewerz:2016zsx}
C.~Ewerz, O.~Kaczmarek and A.~Samberg,
%``Free Energy of a Heavy Quark-Antiquark Pair in a Thermal Medium from AdS/CFT,''
JHEP \textbf{03}, 088 (2018)
doi:10.1007/JHEP03(2018)088
[arXiv:1605.07181 [hep-th]].
%19 citations counted in INSPIRE as of 29 Aug 2021


%\cite{Chen:2017lsf}
\bibitem{Chen:2017lsf}
X.~Chen, S.~Q.~Feng, Y.~F.~Shi and Y.~Zhong,
%``Moving heavy quarkonium entropy, effective string tension, and the QCD phase diagram,''
Phys. Rev. D \textbf{97}, no.6, 066015 (2018)
doi:10.1103/PhysRevD.97.066015
[arXiv:1710.00465 [hep-ph]].
%14 citations counted in INSPIRE as of 29 Aug 2021

%\cite{Arefeva:2018hyo}
\bibitem{Arefeva:2018hyo}
I.~Aref'eva and K.~Rannu,
%``Holographic Anisotropic Background with Confinement-Deconfinement Phase Transition,''
JHEP \textbf{05}, 206 (2018)
doi:10.1007/JHEP05(2018)206
[arXiv:1802.05652 [hep-th]].
%43 citations counted in INSPIRE as of 29 Aug 2021


%\cite{Chen:2018vty}
\bibitem{Chen:2018vty}
X.~Chen, D.~Li and M.~Huang,
%``Criticality of QCD in a holographic QCD model with critical end point,''
Chin. Phys. C \textbf{43}, no.2, 023105 (2019)
doi:10.1088/1674-1137/43/2/023105
[arXiv:1810.02136 [hep-ph]].
%13 citations counted in INSPIRE as of 29 Aug 2021

%\cite{Bohra:2019ebj}
\bibitem{Bohra:2019ebj}
H.~Bohra, D.~Dudal, A.~Hajilou and S.~Mahapatra,
%``Anisotropic string tensions and inversely magnetic catalyzed deconfinement from a dynamical AdS/QCD model,''
Phys. Lett. B \textbf{801}, 135184 (2020)
doi:10.1016/j.physletb.2019.135184
[arXiv:1907.01852 [hep-th]].
%27 citations counted in INSPIRE as of 29 Aug 2021

%\cite{Chen:2019rez}
\bibitem{Chen:2019rez}
X.~Chen, D.~Li, D.~Hou and M.~Huang,
%``Quarkyonic phase from quenched dynamical holographic QCD model,''
JHEP \textbf{03}, 073 (2020)
doi:10.1007/JHEP03(2020)073
[arXiv:1908.02000 [hep-ph]].
%19 citations counted in INSPIRE as of 29 Aug 2021

%\cite{Zhou:2020ssi}
\bibitem{Zhou:2020ssi}
J.~Zhou, X.~Chen, Y.~Q.~Zhao and J.~Ping,
%``Thermodynamics of heavy quarkonium in a magnetic field background,''
Phys. Rev. D \textbf{102}, no.8, 086020 (2020)
doi:10.1103/PhysRevD.102.086020
[arXiv:2006.09062 [hep-ph]].
%11 citations counted in INSPIRE as of 29 Aug 2021


%\cite{Zhou:2021sdy}
\bibitem{Zhou:2021sdy}
J.~Zhou, X.~Chen, Y.~Q.~Zhao and J.~Ping,
%``Thermodynamics of heavy quarkonium in rotating matter from holography,''
Phys. Rev. D \textbf{102}, no.12, 126029 (2021)
doi:10.1103/PhysRevD.102.126029
%1 citations counted in INSPIRE as of 29 Aug 2021

%\cite{Chen:2020ath}
\bibitem{Chen:2020ath}
X.~Chen, L.~Zhang, D.~Li, D.~Hou and M.~Huang,
%``Gluodynamics and deconfinement phase transition under rotation from holography,''
[arXiv:2010.14478 [hep-ph]].
%11 citations counted in INSPIRE as of 29 Aug 2021

%\cite{Chen:2021gop}
\bibitem{Chen:2021gop}
X.~Chen, L.~Zhang and D.~Hou,
%``Running coupling constant at finite chemical potential and magnetic field from holography,''
[arXiv:2108.03840 [hep-ph]].
%1 citations counted in INSPIRE as of 29 Aug 2021


%\cite{LHCb:2017iph}
\bibitem{LHCb:2017iph}
R.~Aaij \textit{et al.} [LHCb],
%``Observation of the doubly charmed baryon $\Xi_{cc}^{++}$,''
Phys. Rev. Lett. \textbf{119}, no.11, 112001 (2017)
doi:10.1103/PhysRevLett.119.112001
[arXiv:1707.01621 [hep-ex]].
%375 citations counted in INSPIRE as of 29 Aug 2021

%\cite{LHCb:2018pcs}
\bibitem{LHCb:2018pcs}
R.~Aaij \textit{et al.} [LHCb],
%``First Observation of the Doubly Charmed Baryon Decay $\Xi_{cc}^{++}\rightarrow \Xi_{c}^{+}\pi^{+}$,''
Phys. Rev. Lett. \textbf{121}, no.16, 162002 (2018)
doi:10.1103/PhysRevLett.121.162002
[arXiv:1807.01919 [hep-ex]].
%91 citations counted in INSPIRE as of 29 Aug 2021

%\cite{Ma:2017nik}
\bibitem{Ma:2017nik}
Y.~L.~Ma and M.~Harada,
%``Chiral partner structure of doubly heavy baryons with heavy quark spin-flavor symmetry,''
J. Phys. G \textbf{45}, no.7, 075006 (2018)
doi:10.1088/1361-6471/aac86e
[arXiv:1709.09746 [hep-ph]].
%26 citations counted in INSPIRE as of 29 Aug 2021


%\cite{Yamamoto:2008jz}
\bibitem{Yamamoto:2008jz}
A.~Yamamoto, H.~Suganuma and H.~Iida,
%``Lattice QCD study of the heavy-heavy-light quark potential,''
Phys. Rev. D \textbf{78}, 014513 (2008)
doi:10.1103/PhysRevD.78.014513
[arXiv:0806.3554 [hep-lat]].
%14 citations counted in INSPIRE as of 29 Aug 2021


%\cite{Najjar:2009da}
\bibitem{Najjar:2009da}
J.~Najjar and G.~Bali,
%``Static-static-light baryonic potentials,''
PoS \textbf{LAT2009}, 089 (2009)
doi:10.22323/1.091.0089
[arXiv:0910.2824 [hep-lat]].
%12 citations counted in INSPIRE as of 29 Aug 2021

%\cite{Andreev:2020xor}
\bibitem{Andreev:2020xor}
O.~Andreev,
%``Some Properties of the $QQq$-Quark Potential in String Models,''
JHEP \textbf{05}, 173 (2021)
doi:10.1007/JHEP05(2021)173
[arXiv:2007.15466 [hep-ph]].
%1 citations counted in INSPIRE as of 29 Aug 2021


%\cite{Andreev:2015iaa}
\bibitem{Andreev:2015iaa}
O.~Andreev,
%``Model of the $N$-quark potential in $SU(N)$ gauge theory using gauge-string duality,''
Phys. Lett. B \textbf{756}, 6-9 (2016)
doi:10.1016/j.physletb.2016.02.070
[arXiv:1505.01067 [hep-ph]].
%8 citations counted in INSPIRE as of 29 Aug 2021

%\cite{Andreev:2015riv}
\bibitem{Andreev:2015riv}
O.~Andreev,
%``Some Aspects of Three-Quark Potentials,''
Phys. Rev. D \textbf{93}, no.10, 105014 (2016)
doi:10.1103/PhysRevD.93.105014
[arXiv:1511.03484 [hep-ph]].
%13 citations counted in INSPIRE as of 28 Aug 2021


%\cite{Andreev:2019cbc}
\bibitem{Andreev:2019cbc}
O.~Andreev,
%``Baryon modes in string breaking from gauge/string duality,''
Phys. Lett. B \textbf{804}, 135406 (2020)
doi:10.1016/j.physletb.2020.135406
[arXiv:1909.12771 [hep-ph]].
%4 citations counted in INSPIRE as of 29 Aug 2021


%\cite{Andreev:2021bfg}
\bibitem{Andreev:2021bfg}
O.~Andreev,
%``Remarks on static three-quark potentials, string breaking and gauge/string duality,''
Phys. Rev. D \textbf{104}, no.2, 026005 (2021)
doi:10.1103/PhysRevD.104.026005
[arXiv:2101.03858 [hep-ph]].
%1 citations counted in INSPIRE as of 29 Aug 2021



%\cite{Yamamoto:2008fm}
\bibitem{Yamamoto:2008fm}
A.~Yamamoto, H.~Suganuma and H.~Iida,
``Heavy-heavy-light quark potential in two approaches,''
Prog. Theor. Phys. Suppl. \textbf{174}, 270-273 (2008).
%doi:10.1143/PTPS.174.270
%[arXiv:0805.4735 [hep-ph]].
%3 citations counted in INSPIRE as of 26 Jan 2022


%\cite{Karch:2006pv}
\bibitem{Karch:2006pv}
A.~Karch, E.~Katz, D.~T.~Son and M.~A.~Stephanov,
%``Linear confinement and AdS/QCD,''
Phys. Rev. D \textbf{74}, 015005 (2006)
doi:10.1103/PhysRevD.74.015005
[arXiv:hep-ph/0602229 [hep-ph]].
%973 citations counted in INSPIRE as of 28 Aug 2021

%\cite{Andreev:2006vy}
\bibitem{Andreev:2006vy}
O.~Andreev,
%``1/q**2 corrections and gauge/string duality,''
Phys. Rev. D \textbf{73}, 107901 (2006)
doi:10.1103/PhysRevD.73.107901
[arXiv:hep-th/0603170 [hep-th]].
%164 citations counted in INSPIRE as of 28 Aug 2021


%\cite{Andreev:2007zv}
\bibitem{Andreev:2007zv}
O.~Andreev,
%``Some Thermodynamic Aspects of Pure Glue, Fuzzy Bags and Gauge/String Duality,''
Phys. Rev. D \textbf{76}, 087702 (2007)
doi:10.1103/PhysRevD.76.087702
[arXiv:0706.3120 [hep-ph]].
%71 citations counted in INSPIRE as of 28 Aug 2021


%\cite{Witten:1998xy}
\bibitem{Witten:1998xy}
E.~Witten,
%``Baryons and branes in anti-de Sitter space,''
JHEP \textbf{07}, 006 (1998)
doi:10.1088/1126-6708/1998/07/006
[arXiv:hep-th/9805112 [hep-th]].
%657 citations counted in INSPIRE as of 28 Aug 2021


%\cite{Andreev:2020pqy}
\bibitem{Andreev:2020pqy}
O.~Andreev,
%``String Breaking, Baryons, Medium, and Gauge/String Duality,''
Phys. Rev. D \textbf{101}, no.10, 106003 (2020)
doi:10.1103/PhysRevD.101.106003
[arXiv:2003.09880 [hep-ph]].
%4 citations counted in INSPIRE as of 28 Aug 2021

\end{thebibliography}
\end{document}